\documentclass[12pt,onecolumn,draftclsnofoot]{IEEEtran}

\usepackage{times}
\usepackage{amsmath, amsthm}
\usepackage{amsfonts}
\usepackage{mathrsfs}
\usepackage{latexsym}
\usepackage{amssymb}
\usepackage{enumerate}
\usepackage{upref}
\usepackage{graphicx}
\usepackage{color}
\usepackage{scrextend}
\usepackage{algorithm}
\usepackage{algpseudocode}
\usepackage{verbatim}
\usepackage{multicol}
\usepackage{colortbl}
\usepackage{multirow}
\usepackage{url}
\usepackage{cite}
\definecolor{mygray}{gray}{.9}

\usepackage{textcomp}
\usepackage[paper=letterpaper,top=1.1in,bottom=0.8in,left=0.8in,right=0.8in]{geometry}

\theoremstyle{definition}

\newtheorem{prop}{Proposition}
\newtheorem{thm}{Theorem}
\newtheorem{cor}{Corollary}
\newtheorem{lem}{Lemma}

\newtheorem{cnstr}[prop]{Construction}

\newtheorem{defn}{Definition}
\newtheorem{exa}{Example}

\newtheorem{rem}{Remark}

\newcommand\bl[1]{{\color{blue}#1}}

\begin{document}

%----------------- The Title Declarations ------------------------------

\title{Storage Codes with Flexible Number of Nodes}
  \author{ 
  Weiqi Li, Zhiying Wang, Taiting Lu, and Hamid Jafarkhani\thanks{The authors are with Center for Pervasive Communications and Computing (CPCC), University of California, Irvine, USA. Emails: \{weiqil4, zhiying, taitingl, hamidj\}@uci.edu. The paper is partially presented in \cite{liflexible}.
} 
}
\maketitle

\begin{abstract}
This paper presents flexible storage codes, a class of error-correcting codes that can recover information from a flexible number of storage nodes. As a result, one can make a better use of the available storage nodes in the presence of unpredictable node failures and reduce the data access latency. Let us assume a storage system encodes $k\ell$ information symbols over a finite field $\mathbb{F}$ into $n$ nodes, each of size $\ell$ symbols. The code is parameterized by a set of tuples $\{(R_j,k_j,\ell_j): 1 \le j \le a\}$, satisfying $k_1\ell_1=k_2\ell_2=...=k_a\ell_a$ and $k_1>k_2>...>k_a = k, \ell_a=\ell$, such that the information symbols can be reconstructed from any $R_j$ nodes, each node accessing $\ell_j$ symbols. In other words, the code allows a flexible number of nodes for decoding to accommodate the variance in the data access time of the nodes. Code constructions are presented for different storage scenarios, including LRC (locally recoverable) codes, PMDS (partial MDS) codes, and MSR (minimum storage regenerating) codes. We analyze the latency of accessing information and perform simulations on Amazon clusters to show the efficiency of presented codes.
%simulations on Amazon cluster show that, compared to fixed codes, the latency can be reduced by up to 6 percent using the flexible storage codes.
\end{abstract}

\section{Introduction}
In distributed systems, error-correcting codes are ubiquitous to achieve high efficiency and reliability. However, most of the codes have a fixed redundancy level, while in practical systems, the number of failures varies over time. When the number of failures is smaller than the designed redundancy level, the redundant storage nodes are not used efficiently. In this paper, we present \emph{flexible storage codes} that make it possible to recover the entire information through accessing a flexible number of nodes.

An $(n,k,\ell)$ \emph{(array) code} over a finite field $\mathbb{F}$ is denoted by $(C_1,C_2,...,C_n), C_i=(C_{1,i}, C_{2,i},\dots,$ $C_{\ell,i})^{T} \in \mathbb{F}^\ell$, where $n$ is the codeword length, $k$ is the dimension, and $\ell$ is the size of each node (or codeword symbol) and is called the \textit{sub-packetization} size. For an $(n,k,\ell)$ code, assume we can recover the entire information by downloading all the symbols from any $R$ nodes. We define the download time of the slowest node among the $R$ nodes as the \emph{data access latency}. In practical systems, the number of available nodes might be different over time and the latency of each node can be modelled as a random variable \cite{liang2014tofec}. Waiting for downloading all $\ell$ symbols from exactly $R$ nodes may result in a large delay. Hence, it is desirable to be able to adjust $R$ and $\ell$ according to the number of failures. Motivated by reducing the data access latency, we propose flexible storage codes below.

%\begin{defn}
A \emph{flexible storage codes} is an $(n,k,\ell)$ code that is parameterized by a given integer $a$ and a set of tuples $\{(R_j, k_j,\ell_j): 1 \le j \le a\}$ that satisfies 
\begin{align}\label{adjustment}
k_j\ell_j=k\ell,1 \le j \le a,k_1>k_2>...>k_a=k, \ell_a = \ell,
\end{align}
and if we take $\ell_j$ particular coordinates of each codeword symbol, denoted by $(C_{m_1,i},C_{m_2,i},\dots,C_{m_{\ell_j},i})^{T} \in \mathbb{F}^{\ell_j}$, $i \in [n]$, where $[n]$ is the set of integers smaller or equal to $n$, we can recover the entire information from any $R_j$ nodes.
%\end{defn}

\begin{figure}[!h]
\centering
\includegraphics[width=\textwidth]{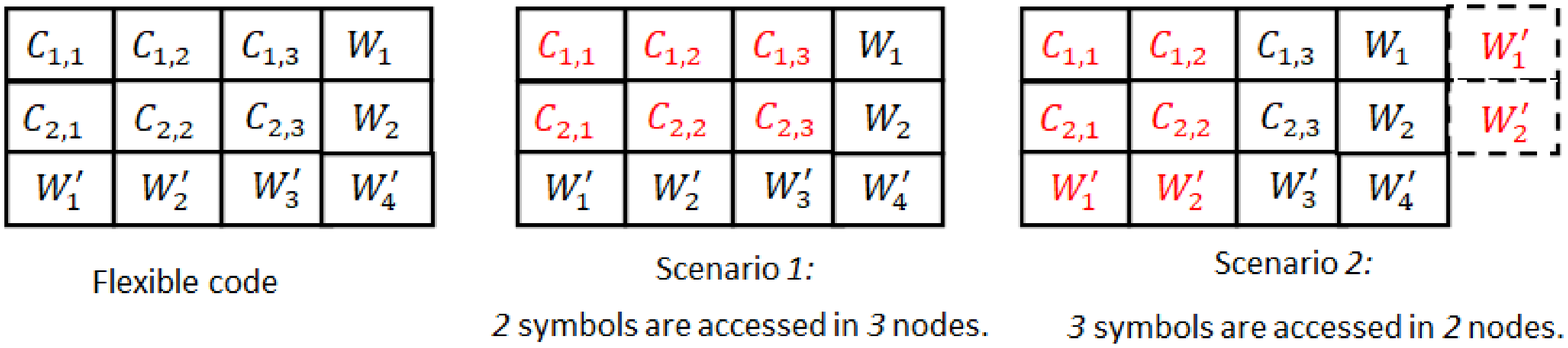}
\caption{Example of a $(4,2,3)$ flexible MDS code. $C_{1,1}, C_{1,2}, C_{1,3}, C_{2,1}, C_{2,2}, C_{2,3}$ are the $6$ information symbols. $W_1=C_{1,1}+C_{1,2}+C_{1,3}, W'_1=C_{1,1}+2C_{1,2}+3C_{1,3}$ are the parities for $ C_{1,1}, C_{1,2},C_{1,3}$, and $W_2= C_{2,1}+ C_{2,2}+C_{2,3}, W'_2=C_{2,1}+2 C_{2,2}+3 C_{2,3}$ are the parities for $C_{2,1}, C_{2,2}, C_{2,3}$. The accessed symbols in each scenario are marked as red. $W'_3 = W'_1 + W'_2, W'_4 = W'_1 + 2 W'_2$ are the parities of $W'_1$ and $W'_2$.
In Scenario 1, all the information symbols are accessed, we obtain the entire information directly. In Scenario 2, $W'_1$ and $W'_2$ are also the parities in Rows $1$ and $2$, respectively. Thus, we obtain $3$ symbols in the first two rows, and the entire information can be decoded.
}
\label{fig: example 1}
\end{figure}

For example, \emph{flexible maximum distance separable (MDS) codes} are codes satisfying the singleton bound for each $k_j$, namely, $R_j = k_j$, $1 \le j \le a$. Fig. \ref{fig: example 1} shows an example. It is easy to see that the flexible code in the example has a better expected latency than a fixed code with either $k=2$ or $3$. In particular, each node can read and then send its three symbols one by one to the decoder (in practice, each symbol can be viewed as, for example, several Megabytes when multiple copies of the same code are applied). The flexible decoder can wait until $2$ symbols from any $3$ nodes, or $3$ symbols from any $2$ nodes are delivered, whose latency is the minimum of the two fixed codes.

Several constructions of flexible MDS codes exist in the literature, though intended for different application scenarios, including error-correcting codes \cite{tamo2019error}, 
universally decodable matrices \cite{ganesan2007existence, ramamoorthy2019universally}, secrete sharing \cite{huang2016communication}, and private information retrieval \cite{bitar2018staircase}. 
However, for other important types of storage codes, such as codes that efficiently recover from a single node failure, or codes that correct mixed types of node and symbol failures, flexible constructions remain an open problem. In this paper, we provide a framework that can produce flexible storage codes for different code families. The main contributions of the paper are summarized below.

{\bf $\bullet$ A framework for flexible codes} is proposed that can generate flexible storage codes given a construction of fixed (non-flexible) storage code.

{\bf $\bullet$ Flexible LRC (locally recoverable) codes} allow information reconstruction from a variable number of available nodes while maintaining the locality property, providing efficient single node recovery. For an $(n,k,\ell,r)$ flexible LRC code parametrized by $\{(R_j,k_j,\ell_j): 1 \le j \le a\}$ that satisfies \eqref{adjustment} and $R_j = k_j+\frac{k_j}{r}-1$, each single node failure can be recovered from a subset of $r$ nodes, while the total information is reconstructed by accessing $\ell_j$ symbols in $R_j$ nodes.
We provide code constructions based on the optimal LRC code construction \cite{tamo2014family}.

{\bf $\bullet$ Flexible PMDS (partial MDS) codes} are designed to tolerate a flexible number of \emph{node failures} and a given number of extra \emph{symbol failures}, desirable for solid-state drives due to the presence of mixed types of failures. We provide an $(n,k,\ell,s)$ with a set of $\{(R_j,k_j,\ell_j): 1 \le j \le a\}$ satisfying \eqref{adjustment} and $R_j = k_j$ such that when $\ell_j$ symbols are accessed in each node, we can tolerate $n-R_j$ failures and $s$ extra symbol failures.
We construct flexible codes from the PMDS code \cite{calis2016general}.

{\bf $\bullet$ Flexible  MSR (minimum storage regenerating) codes} are type of flexible MDS codes such that a single node failure is recovered by downloading the minimum amount of information from the available nodes. Both vector and scalar codes are obtained by applying our flexible code framework to the MSR codes in \cite{ye2017explicit} and \cite{tamo2017optimal}.

{\bf $\bullet$ Latency analysis} is carried out for flexible storage codes. It is demonstrated that our flexible storage codes always have a lower latency compared to the corresponding fixed codes. Also, applying our flexible codes to the matrix-vector multiplication scenario, we show simulation results from Amazon clusters that we can improve $6\%$ for $n=8, R_1=5, R_2=4, \ell_1= 12, \ell_2=15$ and matrix size of $1500 \times 1500$.

{\bf Related work.} The flexibility idea was first proposed in \cite{jafarkhani2019cost} to minimize a cost function such as a linear combination of bandwidth, delay or the number of hops.
Flexible MDS codes were first proposed in \cite{tamo2019error}. In \cite{tamo2019error}, one can recover the entire information by downloading $\ell_j$ symbols from any $k_j$ nodes. However, each of the $k_j$ nodes needs to first read \emph{all} the $\ell$ symbols and then calculate and transmit the $\ell_j$ symbols required for decoding. The aim of \cite{tamo2019error} is to reduce the  bandwidth instead of the number of accessed symbols. 
%The optimal repair code construction is also considered in their work. 
Universally decodable matrices (UDM) \cite{ganesan2007existence,ramamoorthy2019universally} can also be used for the flexible MDS problem.  UDM is a generalization of flexible MDS code where the decoder can obtain different number of symbols from the nodes. In particular, from the first $v_i$ symbols from node $C_i$, for any $v_i, 1 \le i \le n$ such that $\sum\limits_{i=1}^n v_i \geq k\ell$, the entire information can be recovered. Flexibility problems are also considered for secret sharing \cite{huang2016communication, wang2008secret, zhang2012threshold, rawat2018centralized} and private information retrieval \cite{bitar2018staircase, sun2017capacity, beimel2007robust, devet2012optimally, tajeddine2017robust, tajeddine2018robust}, such that the number of available nodes is flexible. The constructions in \cite{huang2016communication} and \cite{bitar2018staircase} are equivalent to each other and they achieved optimal decoding bandwidth while keeping secrecy or privacy from other parties. When we remove the secrecy or privacy requirement, these constructions become flexible MDS codes. All of \cite{tamo2019error,ganesan2007existence,ramamoorthy2019universally, huang2016communication, bitar2018staircase} achieve the optimal field size of $|\mathbb{F}| = n$. 

%While our proposed flexible MDS scheme requires a slightly higher field size $|\mathbb{F}| = n+k_1-k_a$, our general framework allows the generalization to other types of codes such as LRC, PMDS, etc.
%{?? Did we miss any papers on PIR?}
%References other than \cite{huang2016communication} and \cite{bitar2018staircase} do not work for the entire parameters.

%\bl{Summarize the contribution.. UMD ..the number of available nodes and downloaded symbols from each node are both varying .. Communication efficient secrete sharing.. considers the secrete sharing scheme where the number of available nodes is flexible. When there are no secrete constraint, it reduces to flexible MDS code. .. Staircase-PIR.. robust private information retrieval such that the number of available nodes are changing.. .. All of \cite{xxx,xxx,xxx} achieves the optimal field size of $|\mathbb{F}| = n$. While our proposed flexible MDS scheme requires a slightly higher field size $|\mathbb{F}| = n+k_1-k_a$, our general framework allows the generalization to other types of codes such as LRC, PMDS, ect.}

%Erasure-correction codes are not only used for data storage, but can also be applied to distributed computing systems.
There are several works on latency and flexibility in the  literature in distributed coded computing \cite{lee2017speeding, peng2020diversity,yang2019coded,woolsey2020heterogeneous}. 
Specifically, fixed MDS codes are well studied \cite{lee2017speeding}, \cite{peng2020diversity}, where the computing task is distributed to $n$ server nodes and the task can be completed with the results from the fastest $k$ nodes. In \cite{lee2017speeding}, \cite{peng2020diversity}, the authors studied the optimal dimension $k$ under exponential latency of each node. 
Moreover, flexible MDS codes are applied to the distributed computing problem in \cite{yang2019coded, ozfatura2020straggler, woolsey2021coded}. However, it is assumed that we know the set of available nodes before we start computing, which is not the case in our setup. 
Patial computation and straggler model is investigated in \cite{ferdinand2018hierarchical, das2018c}, where each node can compute different amout of tasks. In the setting of \cite{ferdinand2018hierarchical}, the stragglers need to provide partial results and hence are viewed as available nodes. As a result, the system failure tolerance level is lower than that of flexible MDS codes. In the constructions of \cite{das2018c}, the amount of required computation is more than that of flexible MDS codes in the worst case.

%\bl{3. Are there other references on flexibility or latency?}
%\bl{1, Please explain what it means we know the latency?? need to be more specific about  \cite{yang2019coded}, \cite{woolsey2020heterogeneous}. 2. What is the flexibility aspect of \cite{dutta2016short, li2017fundamental, dutta2019optimal}??? If there are none, we can remove these references. 3. Are there other references on flexibility or latency?} Our flexible storage codes can also be applied to the above coded computing problems with the advantage of a lower latency.

The paper is organized as follows: In Section \ref{sec:frameworks}, we present the definition and the construction of our flexible storage codes.
We present the flexible LRC, PMDS, and MSR codes in Sections \ref{subsec:LRC}, \ref{subsec:PMDS}, and \ref{sec:repair}, respectively.
In Section \ref{sec:latency}, we analyze the latency of data access using our flexible codes and compare it with those of fixed codes. The conclusion is made in Section \ref{sec:conclusion}.

\emph{Notation.} For any integer $a\ge 1$, $[a]$ denotes the set $\{1,2,\dots,a\}$. For a matrix $A$ over $\mathbb{F}$, let $rank(A)$ denote its rank. For a set of matrices $A_1,A_2,\dots,A_n$ of size $x \times y$, denote $diag(A_1,A_2,\dots,A_n)$ the corresponding diagonal matrix of size $nx \times ny$. For a finite field $\mathbb{F}$, denote by $\mathbb{F}^\ast = \mathbb{F}\backslash\{0\}.$

\section{The framework for flexible codes}\label{frameworks}
\label{sec:frameworks}
In this section, we define flexible storage codes and provide the framework for flexible codes to convert a fixed (non-flexible) code construction into a flexible one. 
For ease of exposition, ideas are illustrated through flexible MDS code examples in this section. Other types of code constructions are shown in Section \ref{sec:constructions}.

%We first give an example of our framework applied  an MDS code, and then provide the definition and the construction of the general flexible storage codes. 

%In our illustrations, the code is represented by an array. A storage node is denoted by a column, and the $i$-th symbols from all the nodes form the $i$-th row. 

First, we define flexible storage codes. In our illustrations, the codeword is represented by an $\ell \times n$ array over $\mathbb{F}$, denoted as
$C  \in \left( \mathbb{F}^\ell \right)^n $, where $n$ is called the code length, and $\ell$ is called the sub-packetization. Each column corresponds to a storage node.
We choose some fixed integers $j \in [a], \ell_j \in [\ell]$, and \emph{recovery thresholds} $R_j \in [n]$. Let the \emph{decoding columns} $\mathcal{R}_j \subseteq [n]$ be a subset of $R_j$ columns, and the \emph{decoding rows} $\mathcal{I}_1,\mathcal{I}_2,\dots,\mathcal{I}_{R_j} \subseteq [\ell]$ be subsets of rows each with size $\ell_j$.
Denote by $C \mid_{\mathcal{R}_j:\mathcal{I}_1,\mathcal{I}_2,\dots,\mathcal{I}_{R_j}}$ the $\ell_j \times R_j$ subarray of $C$ that takes the rows $\mathcal{I}_1$ in the first column of $\mathcal{R}_j$, the rows $\mathcal{I}_2$ in the second column of $\mathcal{R}_j, \dots$, and the rows $\mathcal{I}_k$ in the last column of $\mathcal{R}_j$. The information will be reconstructed from this subarray.
For flexible MDS codes, flexible MSR codes, and flexible PMDS codes, we have 
$$R_j = k_j.$$
{\bf Notation.} For the above types of codes, we simply omit the parameter $R_j$.
 
For flexible LRC codes, we require 
$$R_j=k_j+\frac{k_j}{n-k_j}+1,$$ 
since the minimum distance is lower bounded by $n-k_j- \frac{k_j}{n-k_j} +2$ \cite{gopalan2012locality}.

\begin{defn}\label{defn1}
The $(n,k,\ell)$ flexible storage code is parameterized by $(R_j,k_j,\ell_j)$, $j \in [a]$, for some positive integer $a$, such that $k_j\ell_j=k\ell,1 \le j \le a,k_1>k_2>...>k_a=k, \ell_a = \ell$. 
It encodes $k \ell$ information symbols  over a finite filed $\mathbb{F}$ into $n$ nodes, each with $\ell$ symbols. The code satisfies the following reconstruction condition for all $j \in [a]$:
from any $R_j$ nodes, each node accesses a set of $\ell_j$ symbols, and we can reconstruct all the information symbols, for any $j \in [a]$. That is, the code is defined by 
\begin{itemize}
\item an encoding function $\mathcal{E}: \left( \mathbb{F}^\ell \right)^k \to  \left( \mathbb{F}^\ell \right)^n$, 
\item decoding functions $\mathcal{D}_{\mathcal{R}_j}:  \left( \mathbb{F}^{\ell_j} \right)^{R_j} \to  \left( \mathbb{F}^\ell \right)^k $, for all $\mathcal{R}_j \subseteq [n], |\mathcal{R}_j|=R_j$, and
\item decoding rows $\mathcal{I}_1,\mathcal{I}_2,\dots,\mathcal{I}_{R_j} \subseteq [\ell]$, $|\mathcal{I}_1|=|\mathcal{I}_2|=\dots=|\mathcal{I}_{R_j}|=\ell_j$, which are dependent on the choice of the decoding columns $\mathcal{R}_j$.
\end{itemize}
The functions are chosen such that any information $U \in \left( \mathbb{F}^\ell \right)^k$ can be reconstructed from the nodes in $\mathcal{R}_j$:
 $$\mathcal{D}_{\mathcal{R}_j} \left( \mathcal{E}(U)\mid _{\mathcal{R}_j:\mathcal{I}_1,\mathcal{I}_2,\dots,\mathcal{I}_{R_j}} \right)  = U. $$
\end{defn}

A \emph{flexible MDS code} is defined as a flexible storage code as in Definition \ref{defn1}, such that $R_j=k_j$.
\begin{comment}
\begin{exa}\label{exa1}
We set $(n,k,\ell)=(4,2,3),(k_1,\ell_1)=(3,2),(k_2,\ell_2)=(2,3)$. The code is defined over the finite field $\mathbb{F}=GF(5)=\{0,1,2,3,4\}$. 
We totally have $k\ell=6$ information symbols and we assume they are $C_{1,1}, C_{1,2}, C_{1,3}, C_{2,1}, C_{2,2}, C_{2,3} \in\mathbb{F}$. We set $W_1=C_{1,1}+C_{1,2}+C_{1,3}, W'_1=C_{1,1}+2C_{1,2}+3C_{1,3}$ be the parities for $ C_{1,1}, C_{1,2},C_{1,3}$, and $W_2= C_{2,1}+ C_{2,2}+C_{2,3}, W'_2=C_{2,1}+2 C_{2,2}+3 C_{2,3}$ be the parities for $C_{2,1}, C_{2,2}, C_{2,3}$. 

The construction is shown as below, each column being a node:
\begin{align}\label{eq:exa1}
  \begin{bmatrix}
C_{1,1}& C_{1,2}& C_{1,3}&W_1\\
C_{2,1}& C_{2,2}& C_{2,3}&W_2\\
W'_1&W'_2&W'_1+W'_2&W'_1+2W'_2
\end{bmatrix}.
\end{align}
$W'_1,W'_2$ are called extra parities generated from the first $2$ rows. Two scenarios of the reconstruction of the flexible code are shown in Fig. xxx.
\end{exa}
\end{comment}
We first examine the example in Fig. \ref{fig: example 1}
\begin{lem}
Fig. \ref{fig: example 1} is an $(n,k,\ell)=(4,2,3)$ flexible MDS code parameterized by $(k_j,\ell_j) \in \{(3,2),(2,3)\}$.
\end{lem}
\begin{IEEEproof}
The encoding function is clear. We have encoded $k\ell=6$ information symbols over $\mathbb{F}$ to a code with $n=4,\ell=3,k=2$.

Then, we present the decoding.
From any $k_1=3$ nodes, each node accesses the first $\ell_1=2$ symbols: The first $2$ rows form a single parity-check $(4,3,2)$ MDS code, and thus we can easily get the information symbols from any $3$ out of $4$ symbols in each row.
From any $k_2=2$ nodes, each node accesses all the $\ell_2=3$ symbols: We can first decode $W'_1$ and $W'_2$ in the last row since the last row is a $(4,2,1)$ MDS code. Then, $(C_{1,1}, C_{1,2}, C_{1,3},W_1,W'_1)$ and $(C_{2,1}, C_{2,2}, C_{2,3},W_2,W'_2)$ form two $(5,3,1)$ MDS codes. We can decode all the information symbols from $W'_1, W'_2$ and any $2$ columns of the first $2$ rows.
\end{IEEEproof}

{\bf Code overview.}
The main idea of the general code construction is similar to that of Fig. \ref{fig: example 1}. The construction is based on a set of $(n + k_j - k_a,k_j,\ell_j-\ell_{j-1})$ codes, each code called a \emph{layer}, such that $k_j\ell_j=k\ell, j\in[a]$, $k_1>k_2>...k_a=k,\ell_a=\ell,\ell_0=0$. The first layer is encoded from the original information symbols and other layers are encoded from the ``extra parities''. 
The intuition for the flexible reconstruction is that after accessing symbols from some layers, we can decode the corresponding information symbols, which is in turn extra parity symbols in an upper layer. Therefore, the decoder can afford accessing less codeword symbols in the upper layer, resulting in a smaller recovery threshold.

\begin{table*}[!ht]
\centering
\caption{Construction of multiple-layer codes}\label{tbl:multiple-layer}
\begin{tabular}{|c|c|c|c|c|c|c|c|c|}

\hline
\multicolumn{4}{|c|}{Storage nodes} &      \multicolumn{5}{|c|}{Extra parities}\\
\hline
$C_{1,1}$&$C_{1,2}$&$\cdots$&$C_{1,n}$ &    $C'_{1,1}$ & $\cdots$&$\cdots$ &$\cdots$& $C'_{1,k_1-k_a}$\\
\hline
$C_{2,1}$&$C_{2,2}$&$\cdots$&$C_{2,n}$ &    $C'_{2,1}$ & $\cdots$ & $\cdots$&  $C'_{2,k_2-k_a}$  & \multicolumn{1}{>{\columncolor{mygray}}l |}{  }\\
\hline
$\vdots$&$\vdots$&$\ddots$&$\vdots$ &               $\vdots$ & $\vdots$ & $\vdots$ &  \multicolumn{1}{>{\columncolor{mygray}}l |}{  }  &  \multicolumn{1}{>{\columncolor{mygray}}l |}{  }  \\
\hline
$C_{a-1,1}$&$C_{a-1,2}$&$\cdots$&$C_{a-1,n}$ &  $C'_{a-1,1}$   & $\cdots$ & $C'_{a-1,k_{a-1}-k_a}$  &  \multicolumn{1}{>{\columncolor{mygray}}l |}{  } &  \multicolumn{1}{>{\columncolor{mygray}}l |}{  }\\
\hline
$C_{a,1}$&$C_{a,2}$&$\cdots$&$C_{a,n}$ &   \multicolumn{1}{|>{\columncolor{mygray}}l |}{  }  &  \multicolumn{1}{>{\columncolor{mygray}}l |}{  }  &  \multicolumn{1}{>{\columncolor{mygray}}l |}{  }  &  \multicolumn{1}{>{\columncolor{mygray}}l |}{  }  &   \multicolumn{1}{>{\columncolor{mygray}}l |}{  } \\
\hline

\end{tabular}

\end{table*}

\begin{cnstr}\label{cnstr:multiple-layers}
In Table \ref{tbl:multiple-layer}, we construct  $(n,k,\ell)$ flexible storage codes parameterized by $\{(k_j,\ell_j): 1 \le j \le a\}$, such that $k_j\ell_j=k\ell$, $k_1>k_2>...k_a=k,\ell_a=\ell$.

Each column is a node. Note that only the first $n$ columns under storage nodes are stored, and the extra parities are auxiliary. Set $\ell_0=0$.
We have $a$ layers, and Layer $j,j\in[a],$ is an $(n+k_j-k_a,k_j,\ell_j-\ell_{j-1})$ code
$$[C_{j,1},C_{j,2},\dots,C_{j,n},C'_{j,1},C'_{j,2},\dots,C'_{j,k_j-k_a}],$$
where $C_{j,i} = [C_{j,1,i},C_{j,2,i},...,C_{j,\ell_j-\ell_{j-1},i}]^T \in\mathbb{F}^{\ell_j-\ell_{j-1}}$, $i \in [n]$, are actually stored, and $C'_{j,i} = [C'_{j,1,i},C'_{j,2,i},...,C'_{j,\ell_j-\ell_{j-1},i}]^T \in \mathbb{F}^{\ell_j-\ell_{j-1}}$, $i \in [k_j-k_a],$ are the auxiliary extra parities.
The $(n+k_1-k_a,k_1,\ell_1)$ code in the first layer is encoded from the $k_1\ell_1=k\ell$ information symbols over $\mathbb{F}$, and the $(n+k_j-k_a,k_j,\ell_j-\ell_{j-1})$ code in Layer $j,j\geq 2,$ is encoded from extra parities $C'_{j',i},$ for $j' \in [j-1],k_j-k_a+1\leq i \leq k_{j-1}-k_a$. As a sanity check, $\sum_{j'=1}^{j-1}(k_{j-1}-k_{j})(\ell_{j'}-\ell_{j'-1})
=(k_{j-1}-k_{j})(\ell_{j-1}-\ell_0 )
=k_j(\ell_j-\ell_{j-1})$ extra parities over $\mathbb{F}$ are encoded into Layer $j$, which matches the code dimension of that layer.
Here we used $\ell_0=0$, and $k_{j-1}\ell_{j-1}=k_j\ell_j$. 
\end{cnstr}

Construction \ref{cnstr:multiple-layers} can be applied to different kinds of codes. We start with MDS codes to show how to use Construction \ref{cnstr:multiple-layers} with a family of storage codes. For an $(n,k,\ell)$ \emph{flexible MDS code} parametrized by $\{(R_j, k_j,\ell_j): 1 \le j \le a\}$ satisfying Definition \ref{defn1}, we have $R_j=k_j$. That is, we can recover the entire information from any $k_j$ nodes, each node accessing its first $\ell_j$ symbols.

\begin{thm}\label{MDS flexible}
With a set of $(n+k_j-k_a,k_j,\ell_j-\ell_{j-1}),j\in[a],\ell_0=0$ MDS codes over $\mathbb{F}$, Construction \ref{cnstr:multiple-layers} is an $(n,k,\ell)$ flexible MDS code parametrized by $\{(R_j,k_j,\ell_j): 1 \le j \le a\}$ satisfying Definition \ref{defn1} and $R_j=k_j$.
\end{thm}

\begin{IEEEproof}
Encoding: As described in Construction \ref{cnstr:multiple-layers}, we encode the $k\ell$ information symbols into an $(n+k_1-k_a,k_1,\ell_1)$ MDS code, and $(n+k_j-k_a,k_j,\ell_j-\ell_{j-1}),2\leq j\leq a$ MDS codes are encoded from the extra parities. 

Decoding: 
%all the information symbols are encoded in in the $(n+k_1-k_a,k_1,\ell_1)$ MDS code in layer $1$. Hence we can decode all the information symbols if we can decode layer 1, i.e., if we get $k_1$ codeword symbols over $\mathbb{F}^{\ell_1}$ in layer $1$.
Fix $j \in [a]$. Assume from any $k_{j}$ nodes, each node accesses its first $\ell_{j}$ symbols over $\mathbb{F}$. We want to show that all the information symbols can be recovered.

We prove by induction that we are able to decode Layer $1$, which contains all the information symbols.

\textbf{Base case}: For Layer $j$, it is obvious since Layer $j$ is an MDS code with dimension $k_j$.

\textbf{Induction step}: Suppose that Layers $j'+1, j'+2,...,j$ are decoded. Then, for Layer $j'$, as shown in Construction \ref{cnstr:multiple-layers} from the decoded layers we get the $k_{j'}-k_j$ extra parities $C'_{j',i},k_{j}-k_a+1\leq i\leq k_{j'}-k_a$. Together with the $k_j$ nodes we have accessed in Layer $j'$, we get enough dimensions to decode Layer $j'$.
\end{IEEEproof}

We note that one can choose any family of MDS codes for the above theorem, e.g., Reed-Solomon codes \cite{reed1960polynomial}, and vector codes \cite{blaum1996mds}. In the case of vector codes, the codeword symbols of the MDS codes are from a vector space rather than a finite field.

\section{Constructions} \label{sec:constructions}
In this section, we show how to apply Construction \ref{cnstr:multiple-layers} to LRC (locally recoverable) codes, PMDS (partial maximum distance separable) codes, and MSR (minimum storage regenerating) codes. These codes provide a flexible reconstruction mechanism for the entire information, and either can reduce the single-failure repair cost, i.e., the number of helper nodes and the amount of transmitted information, or can tolerate mixed types of failures. Applications include failure protection in distributed storage systems and in solid-state drives.

\subsection{Flexible LRC}
\label{subsec:LRC}

An $(n,k,\ell,r)$ LRC code is defined as a code with length $n$, dimension $k$, sub-packetization size $\ell$ and locality $r$. Locality here means that for any single node failure or erasure, there exists a group of at most $r$ available nodes (called helpers) such that the failure can be recovered from them \cite{gopalan2012locality,forbes2014locality, gopalan2014explicit,  zhang2017modified, blaum2020multiple}. The {minimum Hamming distance} of an $(n,k,\ell,r)$ LRC code is lower bounded in \cite{gopalan2012locality} as 
\begin{align}\label{eq:LRC_d}
d_{\min} \ge n-k- \lceil \frac{k}{r} \rceil +2,
\end{align}
and LRC codes achieving the bound are called optimal LRC codes.
For simplicity, we use $(n,k,r)$ LRC codes to present $(n,k,\ell,r)$ LRC codes with $\ell = 1$.
Tamo and Barg \cite{tamo2014family} constructed a family of optimal $(n,k,r)$ LRC codes that encode the $k$ information symbols into $C=[C_{1,1},C_{1,2},\dots,C_{1,r+1},...,$ $C_{\frac{n}{r+1},1},C_{\frac{n}{r+1},2},...,C_{\frac{n}{r+1},r+1}]$, where each group $\{C_{m,i}:i\in[r+1]\},m\in[\frac{n}{r+1}],$ is an MDS code with dimension $r$ and the whole code $C$ has a minimum distance of $n-k- \frac{k}{r}  +2$, i.e., we can decode all the information symbols from any $k+\frac{k}{r}-1$ nodes. 
If an optimal LRC code has the above structure with groups, we say it is an optimal LRC code \emph{by groups}.

We define the \emph{$(n,k,\ell,r)$ flexible LRC code} parameterized by $\{(R_j,k_j,\ell_j):1\le j \le a\}$ as a flexible storage code as in Definition \ref{defn1}, such that all the symbols of any node can be recovered by reading at most $r$ other nodes, and
$$R_j=k_j+\frac{k_j}{n-k_j}+1.$$ 
The above $R_j$ matches the minimum distance lower bound \eqref{eq:LRC_d}. As a result, our definition of flexible LRC code implies optimal minimum Hamming distance when we consider all symbols at each node. 

%\bl{?? Add a picture to explain the definition and the good properties of our construction. Details in Theorem \ref{thm:LRC} and Example \ref{exa:LRC}.}

{\bf Code overview.} The flexible LRC code is based on Construction \ref{cnstr:multiple-layers}, where, first, \emph{extra groups} are generated in each row. Then, $r$ extra parities are chosen from each extra group and encoded into lower layers. During information reconstruction, extra parities and hence extra groups are recovered from lower layers, leading to a smaller number of required access.

\begin{exa}\label{exa: LRC intuition}
Table \ref{tbl:flexible LRC} shows an example of $(n=12,k=4,\ell=3,r=2)$ flexible LRC code. In this code, Rows $1$ and $2$ are $(n=12,k=6,r=2)$ LRC codes encoded from the information, and $1$ extra group is generated in each row. We take $4$ extra parities from the extra groups, which are encoded into $(n=12,k=4,r=2)$ LRC code in Row $3$.
In this example, we have $12$ nodes and they are evenly divided into $4$ groups. Any single failed node can be recovered from the other $2$ nodes in the same group. To recover the entire information, we require either any $8$ nodes, each accessing the first $2$ symbols, or any $5$ nodes, each accessing all $3$ symbols.
The details of this code are shown in Theorem \ref{thm:LRC} and Example \ref{exa:LRC}.
\begin{table*}[!ht]
\centering
\caption{Construction of $(n=12,k=4,\ell=3,r=2)$ flexible LRC code}\label{tbl:flexible LRC}
\begin{tabular}{|c|c|c|c|c|c|c|c|}

\hline
  &\multicolumn{3}{|c|}{group 1} &  $\cdots$&    \multicolumn{3}{|c|}{group 4}\\
\hline
\multirow{2}*{Layer 1}  &$C_{1,1,1}$&$C_{1,1,2}$&$C_{1,1,3}$ &  $\cdots$&  $C_{1,1,10}$ & $C_{1,1,11}$ & $C_{1,1,12}$\\
\cline{2-8}
~ &$C_{1,2,1}$&$C_{1,2,2}$&$C_{1,2,3}$ &  $\cdots$&  $C_{1,2,10}$ & $C_{1,2,11}$ & $C_{1,2,12}$\\
\hline
Layer 2  &$C_{2,1,1}$&$C_{2,1,2}$&$C_{2,1,3}$ &  $\cdots$&  $C_{2,1,10}$ & $C_{2,1,11}$ & $C_{2,1,12}$\\
\hline

\end{tabular}

\end{table*}

\end{exa}

In the following, we apply the optimal LRC codes by groups to Construction \ref{cnstr:multiple-layers} and show how to construct an $(n,k,\ell,r)$ \emph{flexible LRC code} parametrized by $\{(R_j,k_j,\ell_j): 1 \le j \le a\}$ satisfying Definition \ref{defn1}. We assume $n$ is divisible by $r+1$ and all $k_j$'s are divisible by $r$ here. The code is defined in $\mathbb{F}$ of size at least $n+(k_1-k_a)\frac{r+1}{r}$. The resulting code turns out to be an $(n,k_j,\ell_j,r)$ LRC code when $\ell_j$ symbols are accessed at each node. That is, for any single node failure, there exists a group of at most $r$ helpers such that the failure can be recovered from them. 

%\bl{?? what if $k_j-k_a$ is not a multiple of $r$?}

\begin{thm}\label{thm:LRC}
Let $n$ be divisible by $r+1$ and all $k_j, j \in [a]$ be divisible by $r$.
With a set of $(n+(k_j-k_a)\frac{r+1}{r},k_j, r),j\in[a],\ell_0=0$ optimal LRC codes by groups over $\mathbb{F}$, Construction \ref{cnstr:multiple-layers} results in the flexible LRC codes with locality $r$ and $\{(R_j,k_j,\ell_j): 1 \le j \le a\}$ satisfying Definition \ref{defn1}.
\end{thm}

\begin{IEEEproof}
Encoding: In Layer $j$, we apply an $(n+(k_j-k_a)\frac{r+1}{r},k_j, r),j\in[a],\ell_0=0$ optimal LRC code to each row.
As described in Construction \ref{cnstr:multiple-layers}, we encode the $k\ell$ information symbols in the $\ell_1$ rows of Layer $1$, and the remaining rows are encoded from the extra parities.

Next, we show how to choose the $n$ stored symbols and the $k_j-k_a$ extra parities in each row. In the $(n+(k_j-k_a)\frac{r+1}{r},k_j, r)$ LRC code, we have $\frac{n}{r+1}+\frac{k_j-k_a}{r}$ groups. We first pick $\frac{n}{r+1}$ groups, containing $n$ symbols, as the stored symbols. Thus, the $n$ stored symbols in each row form an $(n,k_j, r),j\in[a]$ optimal LRC code.
Then, in the remaining $\frac{k_j-k_a}{r}$ groups, we pick $r$ nodes in each group, which contains $k_j-k_a$ nodes, as the extra parities.

Decoding: Since all the information symbols are encoded in Layer $1$, we can decode the information symbols if we get enough dimensions to decode Layer $1$.

We prove by induction that we can decode all information symbols from any $R_j = k_j+\frac{k_j}{r}-1,j\in[a]$ nodes, each node accesses the first $\ell_j$ symbols.

\textbf{Base case}: From Layer $j$, since each row of it is part of the $(n+(k_j-k_a)\frac{r+1}{r},k_j, r)$ optimal LRC code, we can decode this layer from $R_j$ nodes by the property of the optimal LRC codes.

\textbf{Induction step}: Let $1< j'\leq  j$ be given and suppose that Layers $j', j'+1,...,j$ are decoded. From Construction \ref{cnstr:multiple-layers}, we know that all the extra parities in Layer $j'-1$ are included as the information symbols in Layers $j', j'+1,...,j$ and are decoded. Also, we know from the encoding part that the extra parities in Layer $j'-1$ consist of the $r$ parity symbols in each group of the $(n+(k_{j'-1}-k_a)\frac{r+1}{r},k_{j'-1}, r)$ optimal LRC codes. Thus, according to the locality, the remaining symbol in all $\frac{k_{j'-1}-k_{j}}{r}$ groups in each row can be reconstructed. Therefore, we get additional $(k_{j'-1}-k_{j})\frac{r+1}{r}$ symbols in each row of Layer $j'-1$ from the extra parities. Together with the $R_j$ nodes we accessed in each row of Layer $j'-1$, we get $R_{j'-1}$ symbols and, we are able to decode Layer $j'-1$.

Locality: Since each row is encoded as a LRC code with locality $r$, every layer and the entire code also have locality $r$.

The proof is completed.
\end{IEEEproof}

\begin{exa}\label{exa:LRC}
We set $(n,k,l,r) = (12,4,3,2)$, $(R_1,k_1,\ell_1) = (8,6,2), (R_2,k_2,\ell_2) = (5,4,2)$. The code is defined over $\mathbb{F} = GF(2^4) = \{0,1,\alpha,...,\alpha^{14}\}$, where $\alpha$ is a primitive element of the field.
Totally we have $k\ell = 12$ information symbols and we assume they are $u_{1,0}, u_{1,1},...,u_{1,5},u_{2,0}, u_{2,1},...,u_{2,5}$. The example is based on the optimal LRC code constructions in \cite{tamo2014family}.

The construction is shown below, each column is a node with 3 symbols:
\begin{align}
  \begin{bmatrix}
C_{1,1,1}& C_{1,1,2}& \cdots&C_{1,1,12}\\
C_{1,2,1}& C_{1,2,2}& \cdots&C_{1,2,12}\\
C_{2,1,1}& C_{2,1,2}& \cdots&C_{2,1,12}\\
\end{bmatrix},
\end{align}
where every entry in Row $m$ will be constructed as $f_{m}(x)$ for some polynomial $f_{m}(\cdot)$ and some field element $x$ as below, $m=1,2,3$.

The evaluation points are divided into $4$ groups as $x \in A =\{A_1 = \{1,\alpha^5,\alpha^{10}\}, A_2 = \{\alpha, \alpha^6,\alpha^{11} \}, A_3 = \{\alpha^2,\alpha^7,\alpha^{12}\},$ $A_4 = \{\alpha^3, \alpha^8,\alpha^{13} \}\}$. We also set $A_5 = \{\alpha^4,\alpha^9,\alpha^{14}\}$ as the evaluation points group for the extra parities. 

According to \cite{tamo2014family},
we define $g(x) = x^3$, and one can check $g(x)$ is a constant for each group $A_i$, $i\in[5]$.
Then, the first $2$ rows are encoded with 
\begin{align}
f_m(x) = \big(u_{m,0}+u_{m,1}g(x)+u_{m,2}g^2(x)\big) + x\big(u_{m,3}+u_{m,4}g(x)+u_{m,5}g^2(x)\big), m = 1,2.
\end{align}
The last row is encoded with
\begin{align}
f_3(x) = \big(f_1(\alpha^4)+f_1(\alpha^9)g(x)\big) + x\big(f_2(\alpha^4)+f_2(\alpha^9)g(x)\big).
\end{align}

For each group, since $g(x)$ is a constant, 
$f_m(x),m\in[3]$ can be viewed as a polynomial of degree $2$. Any single failure can be recovered from the other $2$ available nodes evaluated by the points in the same group. The locality $r = 2$ is achieved.

Noticing that $f_1(x)$ and $f_2(x)$ are polynomials of degree $7$, all information symbols can be reconstructed from the first $\ell_1=2$ rows of any $R_1=8$ available nodes.
%To decode all the information symbols, we notice that $f_1(x),f_2(x)$ are polynomials of degree $7$. With any $R_1 = 8$ available nodes, we only need to access the first $\ell_1=2$ rows. 

Moreover, $f_3(x)$ has degree $4$. With $R_2 = 5$ available nodes, we can first decode $f_1(\alpha^4),f_1(\alpha^9),$ $f_2(\alpha^4),$ $f_2(\alpha^9)$ in row $3$. Then, $f_1(\alpha^{14}),f_2(\alpha^{14})$ can be decoded due to the locality $r=2$. At last, together with the $5$ other evaluations of $f_1(x)$ and $f_2(x)$ obtained in Rows $1$ and $2$, we are able to decode all information symbols.

\end{exa}

\subsection{Flexible PMDS codes}
\label{subsec:PMDS}

PMDS codes are first introduced in \cite{blaum2013partial} to overcome mixed types of failures in {Redundant Arrays of Independent Disks} (RAID) systems using {Solid-State Drives} (SSDs).
A code consisting of an $\ell\times n$ array is an $(n,k,\ell,s)$ PMDS code if it can tolerate $n-k$ node or column failures and $s$ additional arbitrary symbol failures in the code.% A general construction of PMDS codes is proposed in \cite{calis2016general} for arbitrary parameters. 

Let $\ell_0=0$ and $\{(k_j,\ell_j): 1 \le j \le a\}$ satisfy \eqref{adjustment}.
We define an  $(n,k,\ell,s)$ \emph{flexible PMDS code} parameterized by $\{(k_j,\ell_j): 1 \le j \le a\}$  such that any row in $[\ell_{j-1}+1, \ell_j]$ is an $(n,k_j)$ MDS code, and from the first $\ell_j$ rows, we can reconstruct the entire information if there are up to $n-k_j$ node failures and up to $s$ additional arbitrary symbol failures, $1 \le j \le a$. As mentioned, for PMDS codes, $R_j=k_j$. Note that different from Definition \ref{defn1}, the number of information symbols for a flexible PMDS code is at most $k\ell-s \triangleq K$.

\begin{exa}
Consider the example of a $(5,3,4,2)$ flexible PMDS code with $\{(k_1,\ell_1), (k_2,\ell_2)\} = \{(4,3),(3,4)\}$ in Table \ref{tab:PMDS}. 
If we only have ``$\ast$'' as failures, we can use the first $4$ nodes to decode, each node accessing the first $3$ symbols.  If both ``$\ast$'' and ``$\triangle$'' are failures, we can decode from Nodes $1,3,4$, each node accessing $4$ symbols. In both cases, the remaining  $K=k\ell-s=10$ symbols are independent and sufficient to reconstruct the entire information. The details of the encoding and decoding for this construction are presented in Theorem \ref{thm:PMDS}.
\end{exa}

\begin{table}[!ht]
\centering
\caption{An example of $(5,3,4,2)$ flexible PMDS code with $\{(k_1,\ell_1), (k_2,\ell_2)\} = \{(4,3),(3,4)\}$. }
\label{tab:PMDS}
\begin{tabular}{|c|c|c|c|c|}

\hline
$C_{1,1,1}$ &   $\triangle$ & $C_{1,1,3}$ & $\ast$ & $\ast$ \\
\hline
$C_{1,2,1}$ &   $\triangle$  & $C_{1,2,3}$ & $C_{1,2,4}$ & $\ast$ \\
\hline
$C_{1,3,1}$ &   $\triangle$  & $\ast$ & $C_{1,3,4}$ & $\ast$ \\
\hline
$C_{2,1,1}$ &   $\triangle$  & $C_{2,1,3}$ & $C_{2,1,4}$ & $\ast$ \\
\hline

\end{tabular}

\end{table}

{\bf Code overview.} To tolerate additional symbol failures, the fixed PMDS code in \cite{calis2016general} uses Gabidulin code to encode the information into auxiliary symbols, which are evenly allocated to each row. Then, an MDS code is applied to the auxiliary symbols in each row, ensuring the protection against column failures. Our flexible PMDS code also encodes the information using Gabidulin code into auxiliary symbols, which are allocated to each layer according to $k_j, j \in [a]$. MDS codes with different dimensions are then applied to each row, thus ensuring flexible information reconstruction. 

A general construction of PMDS codes is proposed in \cite{calis2016general} for any $k$ and $s$ using Gabidulin codes. In this section, we first introduce the construction in \cite{calis2016general} and then show how to apply it to flexible PMDS codes.

An $(N,K)$ Gabidulin code over the finite field $\mathbb{F}=GF(q^L),L\geq N$ is defined by the polynomial $f(x)=\sum_{i=0}^{K-1}u_i x^{q^i}$, where $u_i\in\mathbb{F},i=0,1,...,K-1$ is the information symbol. The $N$ codeword symbols are $f(\alpha_{1}),f(\alpha_{2}),\dots,f(\alpha_{N})$ where the $N$ evaluation points $\{\alpha_1,...,\alpha_N\}$ are linearly independent over $GF(q)$. From any $K$ independent evaluation points over $GF(q)$, the information can be recovered. 

In \cite[Construction 1]{calis2016general}, the $(n,k,\ell,s)$ codeword is an $\ell \times n$ matrix over $\mathbb{F} = GF(q^{k \ell})$ shown below:
\begin{align}\label{pmds}
  \begin{bmatrix}
C_{1,1}& C_{1,2}&\cdots& C_{1,n}\\
C_{2,1}& C_{2,2}&\cdots& C_{2,n}\\
\vdots&\vdots&\ddots& \vdots\\
C_{\ell,1}& C_{\ell,2}&\cdots& C_{\ell,n}\\
\end{bmatrix},
\end{align}
where each column is a node. Set $K=\ell k-s$.
Here, $C_{m,i} \in \mathbb{F},m\in[\ell],i\in[k]$ are the $K+s$ codeword symbols from a $(K+s,K)$ Gabidulin code, and for each row $m$, $m \in [\ell]$,
\begin{align}
[C_{m,k+1},...,C_{m,n}] = [C_{m,1},...,C_{m,k}] G_{\text{MDS}},
\end{align}
where $G_{\text{MDS}}$ is the $k \times (n-k)$ encoding matrix of an $(n,k)$ systematic MDS code over $GF(q)$ that generates the parity.

It is proved in \cite[Lemma 2]{calis2016general} that $t_m$ symbols in row $m, m\in[\ell],$ is equivalent to evaluations of $f(x)$ with $\sum\limits_{m=1}^{\ell}\min(t_m,k)$  evaluation points that are linearly independent over $GF(q)$. Thus, with any $n-k$ node failures and $s$ symbol failures, we have $t_m\leq k$ and
\begin{align}
\sum\limits_{m=1}^{\ell}\min(t_m,k) = \sum\limits_{m=1}^{\ell} t_m = \ell k - s = K.
\end{align}
Then, with the $K$ linearly independent evaluations of $f(x)$, we can decode all information symbols.

Next, we show how to construct flexible PMDS codes. Rather than generating extra parities as in Construction \ref{cnstr:multiple-layers}, the main idea here is that we divide our code into multiple layers, and each layer applies a construction similar to that of \eqref{pmds} with a different dimension.

\begin{thm} \label{thm:PMDS}
We can construct an $(n,k,\ell,s)$ flexible PMDS code over $GF(q^N)$ parameterized by $\{(k_j,\ell_j): 1 \le j \le a\}$ satisfying \eqref{adjustment}, %such that we can tolerate $n-k_j$ node failures and $s$ symbol failures when accessing $\ell_j$ symbols in each node.
with an $(N,K)$ Gabidulin code over $GF(q^N)$, $N=\sum\limits_{j=1}^{a}k_j(\ell_j-\ell_{j-1})$, $K=\ell k -s$, and a set of $(n,k_j)$ systematic MDS codes over $GF(q)$. 
\end{thm}

\begin{IEEEproof}
Encoding: 
Denote $C_{j,m_j,i}$ the symbol in the $m_j$-th row of Layer $j$, and in the $i$-th node, $j\in[a],m_j\in[\ell_j-\ell_{j-1}],i\in[n]$.
We first encode the $K$ information symbols using the $(N,K)$ Gabidulin code. 
Then, we set the first $k_j$ codeword symbols in each row: $C_{j,m_j,i}, j\in[a],m_j\in[\ell_j-\ell_{j-1}],i\in[k_j],$ as the codeword symbols in the  $(N,K)$ Gabidulin code. The remaining $n-k_j$ codeword symbols in each row are 
\begin{align*}
[C_{j,m_j,k_j+1},...,C_{j,m_j,n}] = [C_{j,m_j,1},...,C_{j,m_j,k_j}] G_{n,k_j},
\end{align*}
where $G_{n,k_j}$ is the encoding matrix (to generate the parity check symbols) of the $(n,k_j)$ systematic MDS code over $GF(q)$.

Decoding: For $n-k_J$ failures, we access the first $\ell_J$ rows (the first $J$ layers) from each node. The code structure in each layer is similar to the general PMDS code in \cite[Construction 1]{calis2016general}, from \cite[Lemma 2]{calis2016general} we know that for a union of $t_{m_j}$ symbols in Row $m_j$ of Layer $j$, $j \le J$, they are equivalent to evaluations of $f(x)$ with $\sum\limits_{j=1}^{J}\sum\limits_{m_j=1}^{\ell_j-\ell_{j-1}}\min(t_{m_j},k_j)$ linearly independent points over $GF(q)$ in $GF(q^N)$. 
Thus, with $n-k_J$ node failures and $s$ symbol failures, we have $t_{m_j} \leq k_J \leq k_j$ for $j\in[J]$, and 
\begin{align}
\sum\limits_{j=1}^{J}\sum\limits_{m_j=1}^{\ell_j-\ell_{j-1}}\min(t_{m_j},k_j) = \sum\limits_{j=1}^{J}\sum\limits_{m_j=1}^{\ell_j-\ell_{j-1}}t_{m_j}  = \ell_J k_J - s = K. \nonumber
\end{align}
Then, the information symbols can be decoded from $K$ linearly independent evaluations of $f(x)$.
\end{IEEEproof}

\subsection{Flexible MSR codes}
\label{sec:repair}
In this section, we study flexible MSR codes. In the following, the number of parity nodes is denoted by $r=n-k$ \footnote{Notice that $r$ was used for a different meaning (locality) in LRC codes.}. The \emph{repair bandwidth} is defined as the amount of transmission required to repair a single node erasure, or failure, from all remaining nodes (called helper nodes), normalized by the size of the node. For an $(n,k)$ MDS code, the repair bandwidth is bounded by the minimum storage regenerating (MSR) bound \cite{dimakis2010network} as 
\begin{align}\label{MSR bound}
b\geq \frac{n-1}{n-k}.
\end{align}  
An MDS code achieving the MSR bound is called an MSR code. MSR vector codes are well studied in \cite{ye2017explicit, rashmi2011optimal, papailiopoulos2013repair, Zigzag_Codes_IT, wang2014explicit, rawat2016progress, goparaju2017minimum, ye2016nearly}, where each symbol is a vector. As one of the most popular codes in practical systems, Reed-Solomon (RS) code and its repair is studied in \cite{tamo2017optimal, guruswami2017repairing, dau2018repairing, ye2017repairing,li2019sub}, where each symbol is a scalar. %In this section, we show that with the MSR structure applied to the vector code \cite{ye2017explicit} and the RS code \cite{tamo2017optimal}, we can construct a flexible storage code with optimal repair.

We have shown in Theorem \ref{MDS flexible} that using a set of MDS codes, Construction \ref{cnstr:multiple-layers} can recover the information symbols by any pair $(k_j,\ell_j)$, which means that for the first $\ell_j$ symbols in each node, our code is an $(n,k_j,\ell_j)$ MDS code. In addition, we require the optimal repair bandwidth property for \emph{flexible MSR codes}. 
A \emph{flexible MSR code} is defined to be a flexible storage code as in Definition \ref{defn1}, such that $R_j=k_j$, and a single node failure is recovered using a repair bandwidth satisfying the MSR bound \eqref{MSR bound}.

{\bf Code overview.} Our codes in this section are similar to Construction \ref{cnstr:multiple-layers}, with additional restrictions on the parity check matrices and the extra parities.
The key point here is that the extra parities and the information symbols in lower layers are exactly the same and they also share the same parity check sub-matrix. 
To repair the failed symbol with smallest bandwidth, the extra parities are viewed as additional helpers and the required information can be obtained \emph{for free} from the repair of the lower layers.

We will first show an illustrating example with 2 layers and then present our constructions based on vector and scalar MSR codes, respectively.

\begin{exa}\label{exa2}
We construct an $(n,k,\ell)=(4,2,3)$ flexible MSR code parameterized by $(k_1,\ell_1)=(3,2)$ and $( k_2,\ell_2)=(2,3)$. The reconstruction of the entire information and the repair bandwidth are proved in Lemma \ref{lem:MSR_example}. 

Let $\mathbb{F}=GF(2^2)=\{0,1,\beta,\beta^2=1+\beta\}$, where $\beta$ is a primitive element of $GF(2^2)$.
Our construction is based on the following $(4,2,2)$ MSR vector code over $\mathbb{F}^2$ with parity check matrix
\begin{align}
H=\begin{bmatrix}
h_{1,1}&h_{1,2}&h_{1,3}&h_{1,4}\\
h_{2,1}&h_{2,2}&h_{2,3}&h_{2,4}
\end{bmatrix}=\begin{bmatrix}
0&1&1&0&1&0&0&0\\
1&1&1&1&0&1&0&0\\
0&1&1&1&0&0&1&0\\
1&0&1&0&0&0&0&1
\end{bmatrix},
\end{align}
where each $h_{i,j}$ is a $2\times 2$ matrix over $\mathbb{F}$. Namely, a codeword symbol $c_i$ is in $\mathbb{F}^2$, $i=1,2,3,4$, and
 the codeword $[c_1^T, c_2^T, c_3^T, c_4^T]^T \in (\mathbb{F}^2)^4$ is in the null space of $H$. 
One can check that it is a $(4,2)$ MDS code, i.e., any two codeword symbols suffice to reconstruct the entire information.
The repair matrix is defined as
\begin{align}\label{example repair matrix}
S_1=\begin{bmatrix}
1&0&0&0\\
0&0&0&1
\end{bmatrix},
S_2=\begin{bmatrix}
1&0&0&0\\
0&0&1&0
\end{bmatrix},
S_3=\begin{bmatrix}
1&0&1&0\\
0&1&1&0
\end{bmatrix},
S_4=\begin{bmatrix}
0&1&1&0\\
0&0&0&1
\end{bmatrix}.
\end{align}
It is easy to check that
\begin{align}\label{rank condition}
 rank \begin{pmatrix}
S_\ast \begin{bmatrix}h_{1,i}\\
h_{2,i}
\end{bmatrix}
\end{pmatrix}
=\left\{
\begin{array}{rcl}
2,i=\ast\\
1,i\neq \ast\\
\end{array}. \right. 
\end{align}
When node $\ast \in \{1,2,3,4\}$ fails, we can repair node $c_\ast$ by equations $S_\ast\times H \times [c_1^T,c_2^T,c_3^T,c_4^T]^T=0$. 
In particular, helper $i$, $i\neq \ast$, transmits 
$$S_\ast \begin{bmatrix}h_{1,i}\\
h_{2,i}
\end{bmatrix} c_i,$$
which is $1$ symbol in $\mathbb{F}$,
achieving an optimal total repair bandwidth of $3$ symbols in $\mathbb{F}$.

For our flexible MSR code, every entry in the code array is a vector in $\mathbb{F}^2$.
The code array is shown as below, each column being a node:
\begin{align}\label{eq:exa2}
  \begin{bmatrix}
C_{1,1,1}& C_{1,1,2}& C_{1,1,3}&C_{1,1,4}\\
C_{1,2,1}&C_{1,2,2}& C_{1,2,3}&C_{1,2,4}\\
C_{2,1,1}& C_{2,1,2}& C_{2,1,3}&C_{2,1,4}
\end{bmatrix}.
\end{align}
The code has $2$ layers, where $C_{1,m_1,i} \in \mathbb{F}^2$ are in Layer $1$ and $C_{2,m_2,i}$ are in Layer $2$ with $m_1=1,2,m_2=1,i\in[4]$. Each $C_{j,m_j,i}$ is the vector $[c_{j,m_j,i,1},c_{j,m_j,i,2}]^T$ with elements in $\mathbb{F}$. The code totally contains $48$ bits with $24$ information bits, and each node contains $12$ bits. We define the code with the  $3$ parity check matrices shown below.
Let
\begin{align}\label{eq:msr example parity 1}
H_1=\begin{bmatrix}
h_{1,1}&h_{1,2}&h_{1,3}&h_{1,4}&h_{1,1}\\
h_{2,1}&h_{2,2}&h_{2,3}&h_{2,4}&\beta h_{2,1}
\end{bmatrix},
\end{align}
\begin{align}\label{eq:msr example parity 2}
H_2=\begin{bmatrix}
h_{1,1}&h_{1,2}&h_{1,3}&h_{1,4}&h_{1,2}\\
h_{2,1}&h_{2,2}&h_{2,3}&h_{2,4}&\beta h_{2,2}
\end{bmatrix},
\end{align}
\begin{align}
H_3=\begin{bmatrix}
h_{1,1}&h_{1,2}&h_{1,3}&h_{1,4}\\
\beta h_{2,1}&\beta h_{2,2}&h_{2,3}&h_{2,4}
\end{bmatrix}.
\end{align}
The code is defined by 
\begin{align}
H_1\times [C_{1,1,1}^T,C_{1,1,2}^T,C_{1,1,3}^T,C_{1,1,4}^T,C_{2,1,1}^T]^T =0,\\
H_2\times [C_{1,2,1}^T,C_{1,2,2}^T,C_{1,2,3}^T,C_{1,2,4}^T,C_{2,1,2}^T]^T =0,\\
H_3\times [C_{2,1,1}^T,C_{2,1,2}^T,C_{2,1,3}^T,C_{2,1,4}^T]^T =0.
\end{align}

\end{exa}

\begin{lem} \label{lem:MSR_example}
Example \ref{exa2}  is an $(n,k,\ell)=(4,2,3)$ flexible MSR code parameterized by $(k_j,\ell_j) \in \{(3,2),(2,3)\}$. %And when one node failed, it has the optimal repair bandwidth.
\end{lem}

\begin{IEEEproof}
It is easy to check that the code defined by $H_1$ or $H_2$ is an $(5,2)$ MDS code, and $H_3$ defines an $(4,2)$ MDS code. Thus, the construction in Example \ref{exa2} is the same as Construction \ref{cnstr:multiple-layers}, and the flexible reconstruction of the entire information is shown in Theorem \ref{MDS flexible}.

Let $\ast \in \{1,2,3,4\}$ be the index of the failed node.
For the repair, we first note that
\begin{align}\label{rank condition 2}
 rank \begin{pmatrix}
S_\ast \begin{bmatrix}h_{1,i}\\
h_{2,i}
\end{bmatrix}
\end{pmatrix}
= rank \begin{pmatrix}
S_\ast \begin{bmatrix}h_{1,i}\\
\beta h_{2,i}
\end{bmatrix}
\end{pmatrix}
=\left\{
\begin{array}{rcl}
2,i=\ast\\
1,i\neq \ast\\
\end{array}. \right. 
\end{align}
for $i=1,2$.

Then, we use the same repair matrix $S_\ast$ in \eqref{example repair matrix} to repair the failed node $\ast$:
\begin{align}
S_\ast\times H_1\times [C_{1,1,1}^T,C_{1,1,2}^T,C_{1,1,3}^T,C_{1,1,4}^T,C_{2,1,1}^T]^T =0, \label{example equation 1}\\
S_\ast\times H_2\times [C_{1,2,1}^T,C_{1,2,2}^T,C_{1,2,3}^T,C_{1,2,4}^T,C_{2,1,2}^T]^T =0, \label{example equation 2}\\
S_\ast\times H_3\times [C_{2,1,1}^T,C_{2,1,2}^T,C_{2,1,3}^T,C_{2,1,4}^T]^T =0. \label{example equation 3}
\end{align}
For helper $i \in [4]$, $ i \neq \ast$, it transmits
\begin{align}
& S_\ast \begin{bmatrix}h_{1,i}\\
h_{2,i}
\end{bmatrix} C_{1,1,i}, \label{eq:transmit1}\\
& S_\ast \begin{bmatrix}h_{1,i}\\
h_{2,i}
\end{bmatrix} C_{1,2,i}, \label{eq:transmit2}\\
& S_\ast \begin{bmatrix}h_{1,i}\\
\overline{\beta} h_{2,i}
\end{bmatrix} C_{2,1,i}, \label{eq:transmit3}
\end{align}
where $\overline{\beta} = \beta$ if $i=1,2$ and $\overline{\beta}=1$ if $i=3,4$.
Note that to repair the failed node, in Eq. \eqref{example equation 1} and \eqref{example equation 2}, we also require $S_\ast \begin{bmatrix}h_{1,1}\\\beta h_{2,1}\end{bmatrix} C_{2,1,1}$ and $S_\ast \begin{bmatrix}h_{1,2}\\\beta h_{2,2}\end{bmatrix} C_{2,1,2}$, which can be either obtained from \eqref{eq:transmit3} or solved from Equation \eqref{example equation 3}. %Therefore, we only calculate the required symbols from $C_{2,1,1}$ and $C_{2,1,2}$ in equation \eqref{example equation 3}.

Then, from \eqref{rank condition} and \eqref{rank condition 2} we have that for any failed node, we only need $1$ symbol from each of the remaining $C_{j,m_j,i}$, which meets the MSR bound.
\end{IEEEproof}

{\bf Remark}. Notice that in this example, we do not require the codes in the first layer defined by \eqref{eq:msr example parity 1} and \eqref{eq:msr example parity 2} to be MSR codes, thus resulting in a smaller field. However, the rank condition \eqref{rank condition 2} guarantees the optimal repair bandwidth for the entire code. 
Also, in our general constructions, we do not require the codes in Layers $1$ to $ a-1$ to be MSR codes. 

%Next, we show the constructions for flexible storage codes with optimal repair bandwidth.
 
In the following, we show that by applying Construction \ref{cnstr:multiple-layers} to the vector  MSR code \cite{ye2017explicit} and the RS MSR code \cite{tamo2017optimal}, we can construct flexible MSR codes. %In this section, we have $R_j = k_j$, for simplicity, we just use $k_j$ in the rest of the section.

\vspace{1em}
\subsubsection{Flexible MSR codes with parity check matrices}
\label{subsec:parity}
Below we present codes defined by parity check matrices similar to Example \ref{exa2}.
We show in Theorem \ref{repair theorem} that with certain choices of the parity check matrices, one obtains a flexible MSR code.

\begin{cnstr}\label{optimal repair}
The code is defined in some $\mathbb{F}^L$ parameterized by $(k_j,\ell_j),j\in[a]$ such that $k_j\ell_j=k\ell$, $k_1>k_2>...k_a=k,\ell_a=\ell$. %$L \ge (n-k)^n, |\mathbb{F}| \ge a(n-k)n$.
We define the parity check matrix for the $m_j$-th row in Layer $j\in[a]$ as:
\begin{align}\label{parity check matrix for general MSR constructions}
H_{j,m_j}=  \begin{bmatrix}
h_{j,m_j,1}&\cdots &h_{j,m_j,n}&g_{j,m_j,1}&\cdots & g_{j,m_j,k_j-k_a} \\
\end{bmatrix},
\end{align}
where each $h_{j,m_j,i},g_{j,m_j,i}$ is an $rL\times L$ matrix with elements in $\mathbb{F}$. The $(n+k_j-k_a,k_j)$ MDS code in the $m_j$-th row of Layer $j$ is defined by
\begin{align}
H_{j,m_j}\times [{C_{j,m_j,1}}^T,{C_{j,m_j,2}}^T,\cdots,{C_{j,m_j,n}}^T,{C'_{j,m_j,1}}^{T},\cdots,{C'_{j,m_j,k_j-k_a}}^{T}]^T =0,
\end{align}
where $C_{j,m_j,i}$ are the stored codeword symbols and $C'_{j,m_j,i}$ are the extra parities. In this construction, when we encode the extra parities into lower layers, we set the codeword symbols and the corresponding parity check matrix entries exactly the same. Specifically, for Layers $j < j' \le a$, we set
\begin{align}
g_{j,x,y} &= h_{j',x',y'}, \\
C'_{j,x,y} &= C_{j',x',y'}. 
\end{align}
Here, for $x\in [l_j-l_{j-1}]$, $ k_{j'}-k_{a} +1\leq y \leq k_{j'-1}-k_{a}$, we have $g_{j,x,y}$ corresponds to $h_{j',x',y'}$ in Layer $j'$, and 
\begin{align}
x' &= \lfloor \frac{x(k_{j'-1}-k_{j'}) +y}{k_{j'}} \rfloor, \label{extra parities assignments1}\\
y' &= (x(k_{j'-1}-k_{j'}) +y) \text{ mod } k_{j'}, \label{extra parities assignments2}
\end{align}
where ``mod'' denotes the modulo operation.
\end{cnstr}

For instance, in Example \ref{exa2}, the $2$ extra parities in Layer $1$ are exactly the same as the first $2$ symbols in Layer $2$ with $C'_{1,1,1} = C_{2,1,1},g_{1,1,1} = h_{2,1,1}$ and $C'_{1,2,1} = C_{2,1,2},g_{1,2,1} = h_{2,1,2}$.

\begin{thm}\label{repair theorem}
Assume the parity check matrices of Construction \ref{optimal repair} in \eqref{parity check matrix for general MSR constructions} satisfy

1). [MDS condition.] The codes defined by \eqref{parity check matrix for general MSR constructions} are $(n+k_j-k_a,k_j)$ MDS codes.

2). [Rank condition.] The same repair matrices $S_\ast,\ast\in[n]$ can be used for every parity check matrix such that
\begin{align}\label{optimal repair condition}
rank (S_\ast h_{j,m_j,i})
=\left\{
\begin{array}{rcl}
L,i=\ast\\
\frac{L}{r},i\neq \ast\\
\end{array}, \right. i\in[n].
\end{align}

Then, the code defined by Construction \ref{optimal repair} is a flexible MSR code.

\end{thm}

\begin{IEEEproof}
1). If the MDS property is satisfied, Construction \ref{optimal repair} is the same as Construction \ref{cnstr:multiple-layers} by defining the MDS codes with parity check matrices. The flexible reconstruction of the entire information is presented in Theorem \ref{MDS flexible}.

2). For repair, assume node $\ast,\ast\in[n]$ is failed. We use the repair matrix $S_\ast$ in each row to repair it:
\begin{align}
S_\ast \times H_{j,m_j}\times [{C_{j,m_j,1}}^T,{C_{j,m_j,2}}^T,\cdots,{C_{j,m_j,n}}^T,{C'_{j,m_j,1}}^{T},\cdots,{C'_{j,m_j,k_j-k_a}}^{T}]^T =0.
\end{align}
Notice that $C'_{j,m_j,1},\cdots,C'_{j,m_j,k_j-k_a}$ are also the information symbols in the lower layers with the same corresponding parity check sub-matrices and can be retrieved from the lower layers. Thus, the failed node can be repaired from $n-1$ helpers.% for repair bandwidth we only calculate them once when they are shown in the first $n$ columns. Thus, 
%Thus we only need to transmit symbols in the first $n$ nodes in each row.

Clearly from \eqref{optimal repair condition}, we only need $L/r$ symbols from each helper and the optimal repair bandwidth is achieved.
\end{IEEEproof}

We will now take Ye and Barg's construction \cite{ye2017explicit} to show how to construct the flexible MSR codes satisfying conditions in Theorem \ref{repair theorem}. The code structure in one row is similar to \cite{guruswami2018epsilon}.

%Ye and Barg's construction \cite{ye2017explicit} is defined by a parity check matrix.
Assume the field size $|\mathbb{E}| > rn$ and $\lambda_{i,j} \in \mathbb{E},i\in[n],j=0,1,...,r-1$ are $rn$ distinct elements.
The parity check matrix for the $(n,k)$ MSR code in \cite{ye2017explicit} can be represented as:
\begin{align}\label{Ye and Barg's parity check}
H=  \begin{bmatrix}
I&I&\cdots&I\\
A_1&A_2&\cdots&A_n\\
\vdots&\vdots&\ddots&\vdots\\
A_1 ^{r-1}&A_2 ^{r-1}&\cdots&A_n ^{r-1}\\
\end{bmatrix},
\end{align}
where $I$ is the $L\times L$ identity matrix and $A_i = \sum\limits_{z=0}^{L-1}\lambda_{i,z_i}e_z {e_z} ^T$. $e_z$ is a vector of length $L=r^n$ with all elements $0$ except the $z$-th element which is equal to $1$. We write the $r$-ary expansion of $z$ as $z=(z_{n}z_{n-1}\dots z_{1})$, where $0\leq z_{i} \leq r-1$ is the $i$-th digit from the right and $z = \sum\limits_{i=0}^{r-1} z_i r^i$. Clearly, $A_i$ is an $L\times L$ diagonal matrix with elements $\lambda_{i,z_i}$. The $L\times rL$ repair matrix $S_\ast,\ast\in[n]$ are also defined in \cite{ye2017explicit} and \cite[Sec. IV-A]{guruswami2018epsilon}:
\begin{align}
S_\ast=\text{Diag}(D_\ast,D_\ast,...,D_\ast)
\end{align}
with $\frac{L}{r}\times L$ matrix $D_\ast$, and it is shown that
\begin{align}\label{eq:rank_S}
rank \begin{pmatrix}
S_\ast \begin{bmatrix}
I\\
A_{i}\\
\vdots\\
A_{i}^{r-1}
\end{bmatrix}
\end{pmatrix}
=rank \begin{pmatrix}
D_\ast\\
D_\ast A_{i}\\
\vdots\\
D_\ast A_{i}^{r-1}
\end{pmatrix}
=\left\{
\begin{array}{rcl}
L,i=\ast\\
\frac{L}{r},i\neq \ast\\
\end{array}. \right. 
\end{align}
Here, for $0 \le x \le r^{n-1}-1, 0 \le y \le r^n-1$,
the $(x,y)$-th entry of $D_\ast$ equals 1 if the $r$-ary expansion of $x$ and $y$ satisfies $(x_{n-1},x_{n-1},\dots,x_1) = (y_{n},y_{n-1},\dots,y_{i+1},y_{i-1},\dots,y_1)$, and otherwise it equals 0.

Consider an extended field $\mathbb{F}$ from $\mathbb{E}$ and denote $\mathbb{F}^* \triangleq \mathbb{F}\backslash \{0\}$, $\mathbb{E}^* \triangleq \mathbb{E}\backslash \{0\}$.
Then $\mathbb{F}^*$ can be partitioned to $t\triangleq \frac{|\mathbb{F}^*|}{|\mathbb{E}^*|}$ cosets: $\{\beta_1\mathbb{E}^*,\beta_2\mathbb{E}^*,...,\beta_t\mathbb{E}^*\}$, for some elements $\beta_1,\beta_2,\dots,\beta_t$ in $\mathbb{F}$ \cite[Lemma 1]{li2019sub}. %Then, we apply \eqref{Ye and Barg's parity check} to Construction \ref{optimal repair} and get
Now, we define for the storage nodes (the first $n$ nodes)
\begin{align}\label{Ye and Barg's extension}
h_{j,m_j,i} = \begin{bmatrix}
I\\
\beta_{j,m_j} A_i \\
\beta_{j,m_j}^2 A_i^2 \\
\vdots \\
\beta_{j,m_j}^{r-1}A_i^{r-1} \\
\end{bmatrix},
\end{align}
where $\beta_{j,m_j}$ is chosen from  $\{\beta_1,\beta_2,\dots,\beta_t\}$. We say $\beta_{j,m_j}$ is the \emph{additional coefficient}. 
Then, the extra parity entries $g_{j,m_j,i}$ can be obtained accordingly from \eqref{extra parities assignments1} and \eqref{extra parities assignments2}. Also, notice that $A_i$ might show in $H_{j,m_j}$ several times since the extra parity matrices are the same as the information symbols in lower layers.
We choose the additional coefficients as below.\\
{\bf Condition 1.} In each $H_{j,m_j}$, the additional coefficients for the same $A_i$ are distinct.

\begin{cor}
With parity check matrices defined by \eqref{Ye and Barg's extension} and Condition 1, Construction \ref{optimal repair} is a flexible MSR code.
\end{cor}

\begin{IEEEproof}
We will prove the construction is flexible MSR using Theorem \ref{repair theorem}, for any given $j \in [a], m_j \in [k_j-k_a]$.

1) [MDS condition.] For the codeword $(c_1^T,c_2^T,...,c_{n+k_j-k_a}^T)$ defined by the parity check matrix $H_{j,m_j}$, we write each codeword symbol as $c_i = (c_{i,1}, c_{i,2},...,c_{i,L})^T$. Since $A_i$ is a diagonal matrix, for any $z=0,1,...,L-1,$ we have
%\bl{add columns, change to small font $\gamma_{1},...,\gamma_{k_j-k_a}$. For $y \in [k_j-k_a]$, denote $\gamma_y \triangleq \lambda_{y',z_{y'}}$, corresponding to $g_{j,m_j,y} = h_{j',x',y'}$, where $x',y'$ are computed from \eqref{extra parities assignments1} \eqref{extra parities assignments2} with $x=m_j$.}
\begin{align}\label{separate parity check matrix}
\begin{bmatrix}
1&\cdots&1&1&\cdots&1&\\
\beta_{j,m_j}\lambda_{1,z_1}&\cdots&\beta_{j,m_j}\lambda_{n,z_n}&\alpha_{1}\gamma_1&\cdots&\alpha_{k_j-k_a}\gamma_{k_j-k_a}&\\
\vdots&\ddots&\vdots&\vdots&\ddots&\vdots\\
(\beta_{j,m_j}\lambda_{1,z_1}) ^{r-1}&\cdots&(\beta_{j,m_j}\lambda_{n,z_n}) ^{r-1}&(\alpha_{1}\gamma_1)^{r-1}&\cdots&(\alpha_{k_j-k_a}\gamma_{k_j-k_a}) ^{r-1}&\\
\end{bmatrix}
\begin{bmatrix}
c_{1,z}\\
c_{2,z}\\
\vdots\\
c_{n+k_j-k_a,z}\\
\end{bmatrix} = 0.
\end{align}
Here, $\beta_{j,m_j},\alpha_1,\alpha_2,...,\alpha_{k_j-k_a}$ are additional coefficients satisfying Condition 1.  For $y \in [k_j-k_a]$, denote $\gamma_y \triangleq \lambda_{y',z_{y'}}$, corresponding to $g_{j,m_j,y} = h_{j',x',y'}$, where $x',y'$ are computed from \eqref{extra parities assignments1} and \eqref{extra parities assignments2} with $x=m_j$.
Next, we show \eqref{separate parity check matrix} corresponds to a Vandermonde matrix, i.e., $(c_{1,z},c_{2,z},...,c_{n+k_j-k_a,z})^T$ forms an $(n+k_j-k_a,k_j)$ Reed-Solomon code.
Consider two entries in the second row of the $r \times (n+k_j-k_a)$ matrix in \eqref{separate parity check matrix}. Notice that each entry is the product of an additional coefficient and a $\lambda$ variable (or a $\gamma$ variable). There are three cases.
 1) If the $\lambda$ or the $\gamma$ values are identical, by Condition 1, their additional coefficients differ. So, these two entries are distinct. 2) If the $\lambda$ or the $\gamma$ values are distinct, and the additional coefficients are identical, then the two entries are distinct. 3) The $\lambda$ or the $\gamma$ values are distinct, and the additional coefficients are distinct. Noticing $\lambda$ and $\gamma$ belong to $\mathbb{E}^*$, distinct additional coefficients implies that the two entries are in distinct cosets.

After we combine all $z =0,1,\dots,L-1$ together, $(c_1^T,c_2^T,...,c_{n+k_j-k_a}^T)^T$ is an $(n+k_j-k_a,k_j)$ MDS vector code. 

2) [Rank condition.]
Multiplying the row of a matrix by a constant does not change the rank. So, by \eqref{eq:rank_S} and \eqref{Ye and Barg's extension},
\begin{align}
rank(S_\ast h_{j,m_j,i})
=rank \begin{pmatrix}
D_\ast\\
D_\ast \beta A_{i}\\
\vdots\\
D_\ast \beta^{r-1}A_{i}^{r-1}
\end{pmatrix}
=rank \begin{pmatrix}
D_\ast\\
D_\ast A_{i}\\
\vdots\\
D_\ast A_{i}^{r-1}
\end{pmatrix} 
%=rank \begin{pmatrix}
%S_\ast \begin{bmatrix}
%I\\
%A_{i}\\
%\vdots\\
%A_{i}^{r-1}
%\end{bmatrix}
%\end{pmatrix}
=\left\{
\begin{array}{rcl}
L,i=\ast\\
\frac{L}{r},i\neq \ast\\
\end{array}. \right. 
\end{align}

Since the code satisfies the above two conditions, using Theorem \ref{repair theorem}, it is a flexible MSR code.
\end{IEEEproof}

%Thus, by applying \eqref{Ye and Barg's extension} we can get the flexible recoverable codes with optimal repair.

To calculate the required field size, we study how many additional coefficients are required for our flexible MSR codes satisfying Condition 1. In the following, we propose $2$ possible coefficient assignments. It should be noticed that one might find better assignments with smaller field sizes. %In each $H_{j,m_j}$ we require distinct $\beta$ in front of the same $A_i$ to guarantee the MDS property. 

%\bl{Assign different $\beta$ to different rows for the storage nodes.}

The simplest coefficient assignment assigns different additional coefficients to different rows, i.e., $\beta_{j,m_j}$ to Row $m_j$ in Layer $j$ for the storage nodes (the first $n$ nodes). By doing so, the parity check matrix $\beta_{j,m_j} A_i, j\in[a],m_j\in[\ell_j-\ell_{j-1}] i\in[n]$ will show at most twice in Construction \ref{optimal repair}, i.e., in Layer $j$ corresponding to storage Node $i$, and in Layer $j'$ corresponding to an extra parity, for some $j>j'$. Hence, the same $A_i$ will correspond to different additional coefficients in the same row and Condition 1 is satisfied. In this case, we need a field size of $\ell|\mathbb{E}|$.

%\bl{Rephrase.}
%\bl{Each row is indexed by a pair $(j,m_j)$.}

In the second assignment, we assign different additional coefficients in different layers for the storage nodes (the first $n$ nodes), but for different rows in the same layer, we might use the same additional coefficient. For a given row, the storage nodes will not conflict with the extra parities since the latter correspond to the storage nodes in other layers. Also, the extra parities will not conflict with each other if they correspond to the storage nodes in different layers. 
Then, we only need to check the extra parities in the same row corresponding to storage nodes in the same layer. For the extra parities/storage nodes $g_{j,x,y} = h_{j',x',y'}$, given $j,x,j',y'$, the additional coefficients should be different. In this case $  k_{j'}-k_{a} +1\leq y \leq k_{j'-1}-k_{a}$, and there will be at most $\lceil \frac{k_{j'-1}-k_{j'}}{k_{j'}} \rceil$ that make $y'$ a constant in \eqref{extra parities assignments2}. As long as we assign $\lceil \frac{k_{j'-1}-k_{j'}}{k_{j'}} \rceil$ number of $\beta$ in Layer $j',j'\ge 2$  (in Layer 1 we only need one $\beta$), Condition 1 is satisfied.

The total number of required additional coefficients is $1 + \sum\limits_{j=2}^a \lceil \frac{k_{j-1}-k_j}{k_j} \rceil \triangleq t$. Notice that $(k_{j-1}-k_j)\ell_{j-1} = k_j(\ell_j-\ell_{j-1})$, we have 
\begin{align}\label{eq:assignment of a}
t =1 + \sum\limits_{j=2}^a \lceil \frac{k_{j-1}-k_j}{k_j} \rceil = 1 +  \sum\limits_{j=2}^a \lceil \frac{\ell_j-\ell_{j-1}}{\ell_{j-1}} \rceil \le 1 +  \sum\limits_{j=2}^a (\ell_j-\ell_{j-1})\le \ell.
\end{align}
Moreover, in the best case when we have  $k_{j-1}-k_j\leq k_j$ for all $j$, the number of additional coefficients is $a$, and $|\mathbb{F}| \ge a |\mathbb{E}|$. %\bl{Is it smaller than $\ell$?}

Here, we briefly compare our construction with another flexible MSR construction in \cite{tamo2019error}.
In our code, each node is in $\mathbb{F}^{\ell(n-k)^n}$, where $|\mathbb{F}|\geq t(n-k)n$. Namely, each node requires $\ell(n-k)^n \log_2(t(n-k)n)$ bits. Tamo, Ye and Barg also considered the optimal repair of flexible codes in \cite{tamo2019error} under their setting, i.e., the downloaded symbols instead of the accessed symbols in each node is flexible to reconstruct the entire information. Their nodes are elements in $\mathbb{F}^{s(n-k)^n}$, and  $|\mathbb{F}| \geq s (n-k) n$,
%They require at least $s(n-k)^n$ symbols of field $|\mathbb{F}| \geq s (n-k) n$ in each node, 
where $s$ is defined such that $ s_j/s = \ell_j/\ell$ fraction of information are downloaded in each node, where $s$ is the least common multiple of $s_1,s_2,...,s_a$. Without loss of generality, we can choose $\ell = s$ in our construction. Hence, for Eq. \eqref{eq:assignment of a}, the required field size of our construction is better than that of the construction in \cite{tamo2019error}.

\subsubsection{Flexible RS MSR codes}
\label{subsec:RS}
In this section, we introduce the construction of Reed-Solomon (RS) MSR codes.

%\bl{Rephrase.}
An $RS(n,k)$ code over the finite field $\mathbb{F}$ is defined as
\begin{align*}
RS(n,k)=\{  (f(\alpha_{1}),f(\alpha_{2}),\dots,f(\alpha_{n})): f\in \mathbb{F}[x], \deg(f)\le k-1 \},
\end{align*}
where the evaluation points are defined as $\{\alpha_{1},\alpha_{2},\dots,\alpha_{n}\}\subseteq \mathbb{F}$, and $\deg()$ denotes the degree of a polynomial. The encoding polynomial $f(x)=u_{0}+u_{1}x+\dots+u_{k-1}x^{k-1}$, where $u_{i} \in \mathbb{F}, i=0,1, \dots,k-1$ are the information symbols. Every evaluation symbol $f(\alpha_i),i\in[n]$ is called a codeword symbol. RS codes are MDS codes, namely, from any $k$ codeword symbols, the information can be recovered. 

Let $\mathbb{B}$ be the base field of $\mathbb{F}$ such that $\mathbb{F}=\mathbb{B}^L$.
For repairing RS codes, \cite{guruswami2017repairing} and \cite{li2019sub} shows that any linear repair scheme for a given $RS(n,k)$ over the finite field $\mathbb{F}=\mathbb{B}^L$ is equivalent to finding a set of repair polynomials $p_{\ast,v}(x)$ such that for the failed node $f(\alpha_\ast)$, $\ast \in [n]$,
\begin{align}
rank_\mathbb{B}(\{p_{\ast,v}(\alpha_\ast):v \in [L]\}) = L ,
\end{align}
where the rank $rank_{\mathbb{B}}(\{\gamma_1,\gamma_2,...,\gamma_i\})$ is defined as  the cardinality of a maximum subset of $\{\gamma_1,\gamma_2,...,\gamma_i\}$ that is linearly independent over $\mathbb{B}$.

The transmission from helper $f(\alpha_i)$ is
\begin{align}\label{RS repair transmission}
Tr_{\mathbb{F}/\mathbb{B}}(p_{\ast,v}(\alpha_i)f(\alpha_i)), v\in[L],
\end{align}
where the trace function $Tr_{\mathbb{F}/\mathbb{B}}(x)$ is a linear function such that for all $x\in\mathbb{F}$, $Tr_{\mathbb{F}/\mathbb{B}}(x)\in\mathbb{B}$ \cite{lidl1994introduction}. The repair bandwidth for the $i$-th helper is
\begin{align}
b_i=rank_\mathbb{B}(\{p_{\ast,v}(\alpha_i):v \in L\})
\end{align}
symbols in $\mathbb{B}$.

The flexible RS MSR code construction is similar to Construction \ref{optimal repair} based on parity check matrices, as presented below.

%\bl{Change...}
\begin{cnstr}\label{RS optimal repair construction}
We define a code in $\mathbb{F}=GF(q^L)$ with a set of pairs $(k_j,\ell_j),j\in[a]$ such that $k_j\ell_j=k\ell$, $k_1>k_2>...k_a=k,\ell_a=\ell$, $r=n-k$.
In the $m_j$-th row in Layer $j\in[a]$, the codeword symbols $C_{j,m_j,i},i\in[n]$ are defined as:
\begin{align}
C_{j,m_j,i} = f_{j,m_j} (\alpha_{j,m_j,i}),
\end{align}
and the extra parities $C'_{j,m_j,i},i\in[k_j-k_a]$ are defined as
\begin{align}
C'_{j,m_j,i} = f_{j,m_j} (\alpha_{j,m_j,i+n}),
\end{align}
where $\{f_{j,m_j}(\alpha_{j,m_j,i}), i\in[n+k_j-k_a]\}$ is an $RS(n+k_j-k_a, k_j)$ code. We next define the encoding polynomial $f_{j,m_j}(x)$ and the evaluation point $\alpha_{j,m_j,i}$.

In this construction, we set the extra parities and the corresponding evaluation points exactly the same as the information symbols in lower layers, and we arrange the extra parities the same way as in Construction \ref{optimal repair}.
Specifically, for $C'_{j,x,y}$ in Layer $j$, $x\in [l_j-l_{j-1}]$, when $ k_{j}-k_{j'-1} +1\leq y \leq k_{j}-k_{j'}$ for $j+1\leq j'\leq a$, it is encoded to Layer $j'$ with $\alpha_{j,x,y+n} = \alpha_{j',x',y'}$ and $C'_{j,x,y} = C_{j',x',y'}$, with $x',y'$ in \eqref{extra parities assignments1} \eqref{extra parities assignments2}. The encoding polynomial $f_{j',m_{j'}}(x)\in\mathbb{F}$ in Layer $j'$ is defined by the $k_{j'}$ evaluation points and the codeword symbols from the extra parities.
\end{cnstr}

\begin{thm}\label{repair thm for RS code}
Construction \ref{RS optimal repair construction} is a flexible MSR RS code, if it satisfies:

1) [MDS condition.] In Row $m_j$ of Layer $j$, $\alpha_{j,m_j,i}, i\in [n+k_j-k_a]$ are distinct elements in $\mathbb{F}$.

2) [Rank conditions.] The same set of repair polynomials $p_{\ast,v}(x),\ast\in[n],v\in [L],$ can be used in each row such that:
\begin{align}\label{repair condition in Thm 1}
rank_\mathbb{B}(\{p_{\ast,v}(\alpha_{j,m_j,\ast}):v \in [L]\}) = L, 
\end{align}
\begin{align}\label{RS thm repair bandwidth}
b_i=rank_\mathbb{B}(\{p_{\ast,v}(\alpha_{j,m_j,i}):v \in [L]\}) = L/r , i\in[n]\backslash \{*\}.
\end{align}
\end{thm}

\begin{IEEEproof}
1). In the case when $\alpha_{j,m_j,i}, i\in [n+k_j-k_a]$ are distinct elements in $\mathbb{F}$, $\{f_{j,m_j}(\alpha_{j,m_j,i}), i\in [n+k_j-k_a]\}$ is $RS(n+k_j-k_a,k_j)$. Moreover, Layer $j'$ is encoded from the $k_{j'}$ extra parities in Layers $1,2,\dots,j'-1$.
Thus,  Construction \ref{RS optimal repair construction} is the same as Construction \ref{cnstr:multiple-layers} by using the RS codes as the MDS codes. The flexible reconstruction property is shown in Theorem \ref{MDS flexible}.

2). For the repair, since the extra parities share the same codeword symbols and evaluation points with the storage nodes in lower layers, from \eqref{RS repair transmission} we know that the transmission for repair is also the same. Thus, we only transmit them once when they are shown as storage nodes. 

From \eqref{RS thm repair bandwidth} we know that in each row, each helper transmits $L/r$ symbols, which is optimal.
\end{IEEEproof}

We take the construction in \cite{li2019sub} as the $RS(n+k_j-k_a,k_j),j\in[a]$ codes in Construction \ref{RS optimal repair construction} to show how to construct flexible MSR RS codes.

In \cite[Theorem 5]{li2019sub}, the RS code is defined in $\mathbb{F}$ with evaluation points chosen from $\{\beta_1\alpha_i,$ $\beta_2\alpha_i,...,\beta_t\alpha_i,i\in[n]\}$ such that $t = \frac{|\mathbb{F^*}|}{|\mathbb{E^*}|}$ for a subfield $\mathbb{E}=GF(q^L)$ of $\mathbb{F}$, and $\alpha_i \in \mathbb{E}$, $i \in [n]$. 
Here $\beta_1,\dots,\beta_t$ correspond to elements in $\mathbb{F}$ such that $\{\beta_1 \mathbb{E}^\ast, \dots, \beta_t \mathbb{E}^\ast\}$ forms a partition of $\mathbb{F}^\ast$ \cite[Lemma 1]{li2019sub}.
For the repair polynomials $p_{\ast,v}(x)$ in \cite{li2019sub},
\begin{align}\label{repair condition for RS ref}
rank_\mathbb{B}(\{p_{\ast,v}(\beta\alpha_i):v \in [L]\}) =\left\{
\begin{array}{rcl}
L,i=\ast\\
\frac{L}{r},i\neq \ast\\
\end{array}\right.
\end{align}
for all $\beta$ chosen from $\{\beta_1,...,\beta_t\}$. The required subfield size in \cite{li2019sub} is $|\mathbb{E}| \approx n^n$.

For Construction \ref{RS optimal repair construction}, we assign the evaluation points in the storage nodes as $\alpha_{j,m_j,i} = \beta_{j,m_j}\alpha_i\in\mathbb{F},$ $i \in [n], j\in[a], m_j \in [\ell_j-\ell_{j-1}]$, where $\beta_{j,m_j}$ is chosen from $\{\beta_1,\dots,\beta_t\}$. The evaluation points of the extra parities are given by the storage nodes as in \eqref{extra parities assignments1} and \eqref{extra parities assignments2}. 
We assign the additional coefficient $\beta$ to satisfy Condition 1. Similar to Construction \ref{optimal repair}, we guarantee that in each row, the $n+k_j-k_a$ evaluation points are distinct and the total number of required $\beta$ required is $t =1+\sum\limits_{j=2}^a \lceil \frac{k_{j-1}-k_j}{k_j} \rceil$. In the best case when we have  $k_{j-1}-k_j\leq k_j$ for all $j$, the number of $\beta$ we required is $a$. The required field size is $a|\mathbb{E}|$.

\begin{cor}
With the RS code in \cite{li2019sub}, Construction \ref{RS optimal repair construction} is a flexible MSR RS code.
\end{cor}

\begin{IEEEproof}
We use Theorem \ref{repair thm for RS code} to prove that the code is a flexible MSR RS code.

1) [MDS condition.] We have assigned the evaluation points in each row as distinct elements in $\mathbb{F}$.

2) [Rank conditions.] We know from \eqref{repair condition for RS ref} that the rank conditions in Theorem \ref{repair thm for RS code} are satisfied.
\end{IEEEproof}

\section{Latency}
\label{sec:latency}

In this section, we analyze the latency of obtaining the entire information using our codes with flexible number of nodes. 
%As analyzed in Section \ref{constructions}, for any pair of $(k_j,\ell_j)$, when $k_j$ nodes have accessed the first $\ell_j$ symbols, we can reconstruct all the information symbols. 
%We note that the latency model in the following discussion mainly captures the time to read and to transmit data from the storage nodes to the decoder. The decoding computation time is assumed to be less significant. 

One of the key properties of the flexible storage codes presented in this paper is that the decoding rows are the first $\ell_j$ rows if we have $R_j$ available nodes. As a result, the decoder can simply download symbols one by one from each node, and symbols of Layer $j$ can be used for Layers $j,j+1,\dots,a$. 

For one pair of $(R_j,\ell_j)$, define a random variable $T_j$ associated with the time for the first $R_j$ nodes transmitting the first $\ell_j$ symbols. $T_j$ is called the \emph{latency} for the $j$-th layer.
Instead of predetermining a fixed pair $(R,\ell)$ for the system, flexible storage codes allow us to use all possible pairs $(R_j,\ell_j), j \in [a]$. The decoder downloads symbols from all $n$ nodes and as long as it obtains $\ell_j$ symbols from $R_j$ nodes, the download is complete. 
For flexible codes with Layers $1,2,...,a$, we use $T_{1,2,...,a}= \min(T_j,j\in[a])$ to represent the \emph{latency}.

Notice that for the fixed code with the same failure tolerance level, i.e., $R=R_a, \ell = \ell_a$, its latency is $T_a$. Since
\begin{align}
T_{1,2,...,a} = \min(T_j,j\in[a])\leq T_a,
\end{align}
we reach the following remark.

\begin{rem} \label{rem:latency}
Given the storage size per node $\ell$, the number of nodes $n$, and recovery threshold $R=R_a$, the flexible storage code can reduce the latency of obtaining the entire information compared to any fixed array code.
\end{rem}

Assume the probability density function (PDF) of $T_j$ is $p_{R_j,\ell_j}(t)$. We calculate the expected delay as
\begin{align}
E(T_j)=\int_0^\infty \tau_j p_{R_j,\ell_j}(\tau_j)d\tau_j.
\end{align}
If a fixed code is adopted, one can optimize the expected latency and get an optimal pair $(R^\ast,\ell^\ast)$ for a given distribution \cite{lee2017speeding}, \cite{peng2020diversity}. However, a flexible storage code still outperforms such an optimal fixed code in latency due to Remark \ref{rem:latency}. Moreover, in practice the choice of $(n,k,R,\ell)$ depends on the system size and the desired failure tolerance level and is not necessarily optimized for latency.

Next, we take the \emph{Hard Disk Drive} (HDD) storage system as an example to calculate the latency of our flexible storage codes and show how much we can save compared to a fixed MDS code. In this part, we compute the overall latency of a flexible code with $(R_1,\ell_1)$, $(R_2,\ell_2)$, and length $n$. We compare it with the latency of fixed codes with $(n,R_1,\ell_1)$ and $(n,R_2,\ell_2)$, respectively.

The HDD latency model is derived in \cite{ruemmler1994introduction}, where the overall latency consists of the \emph{positioning time} and the \emph{data transfer time}. The positioning time measures the latency to move the hard disk arm to the desired cylinder and rotate the desired sector to under the disk head. 
As the accessed physical address for each node is arbitrary, we assume the positioning time is a random variable uniformly distributed, denoted by  $U(0,t_{\text{pos}})$, where $t_{\text{pos}}$ is the maximum latency required to move through the entire disk. The data transfer time is simply a linear function of the data size, and we assume the transfer time for a single symbol in our code is $t_{\text{trans}}$.
Therefore, the overall latency model is $X + \ell \cdot t_{\text{trans}} $, where $X\thicksim U(0,t_{\text{pos}})$ and $\ell$ is the number of accessed symbols.

Consider an $(n,R,\ell)$ fixed code. When $R$ nodes finish the transmission of $\ell$ symbols, we get all the information. The corresponding latency is called the $R$-th order statistics. For $n$ independent random variables satisfying $U(0,t_{\text{pos}})$, the $R$-th order statistics for the positioning time, denoted by $U_R$, satisfies a \emph{beta distribution} \cite{jones2009kumaraswamy}:
\begin{align}
U_{R} \thicksim \text{Beta}(R,n+1-R,0,t_{\text{pos}}).
\end{align}
with expectation $E[U_{R}] = \frac{R}{n+1}t_{\text{pos}}$.
For a random variable $Y \thicksim \text{Beta}(\alpha,\beta,a,c)$, the probability density function (pdf) is defined as
\begin{align}
f(Y=y; \alpha,\beta,a,c) = \frac{(y-a)^{\alpha-1}(c-y)^{\beta-1}}{(c-a)^{\alpha+\beta - 1}B(\alpha,\beta)},
\end{align}
where
\begin{align}
B(\alpha,\beta) = \int_{t=0}^{1}t^{\alpha-1}(1-t)^{\beta-1}d t
\end{align}
is the \emph{Beta function}.

The expectation of overall latency for an $(n,R_1,\ell_1)$ fixed code, denoted by $T_1$, is
\begin{align}\label{Latency of fixed code 1}
E(T_1) = \frac{R_1}{n+1} t_{\text{pos}} +\ell_1 t_{\text{trans}}.
\end{align}
Similarly, the expected overall latency $E(T_2)$ for the fixed $(n,R_2,\ell_2)$ code is
\begin{align}\label{Latency of fixed code 2}
E(T_2) = \frac{R_2}{n+1} t_{\text{pos}} +\ell_2 t_{\text{trans}}.
\end{align}

Now, consider our flexible code with 2 layers. The difference of the positioning times $U_{R_1}$ and $U_{R_2}$ is
\begin{align}
 \Delta U = U_{R_1} - U_{R_2} \thicksim \text{Beta}(R_1-R_2,n+1-(R_1-R_2),0,t_{\text{pos}}).
\end{align}
Thus, we can get the expectation of the overall latency for our flexible code, denoted by $T_{1,2}$, as
\begin{align}
E(T_{1,2}) 
&= E(min(T_1,T_2)) \nonumber\\
&= E(T_1|T_1-T_2 \le 0) P(T_1-T_2 \le 0) +  E(T_2|T_1-T_2>0)P(T_1-T_2>0) \nonumber\\
&= E(T_1) - E(T_1-T_2|T_1-T_2>0) P(T_1-T_2>0) \nonumber\\
&= \frac{R_1}{n+1} t_{\text{pos}} +\ell_1 t_{\text{trans}} - \int_{(\ell_2-\ell_1) t_{\text{trans}}} ^{t_{\text{pos}}} [\Delta U - (\ell_2-\ell_1) t_{\text{trans}}] f(\Delta U) d\Delta U,
\end{align}
where the last term is the saved latency compared to an $(n,R_1,\ell_1)$ code. The saved latency can be calculated as:
\begin{align}
E(T_1 - T_{1,2})  &= \int_{(\ell_2-\ell_1) t_{\text{trans}}} ^{t_{\text{pos}}} [\Delta U - (\ell_2-\ell_1) t_{\text{trans}}] f(\Delta U) d\Delta U \\ \nonumber
&= \frac{at_{\text{pos}}}{a+b}I_{1-x}(b,a+1) - (\ell_2-\ell_1)t_{\text{trans}}I_{1-x}(b,a),
\end{align}
where $x = \frac{\ell_2-\ell_1}{t_{\text{pos}}}t_{\text{trans}}, a=R_1-R_2, b=n-(R_1-R_2) + 1 $, and $I_{x}(a,b)$ is the \emph{regularized incomplete beta function}:
\begin{align}
I_{x}(a,b) = \frac{B(x;a,b)}{B(a,b)},
\end{align}
with \emph{incomplete beta function}
\begin{align}
B(x;a,b) = \int_{t=0}^{x}t^{a-1}(1-t)^{b-1}d t.
\end{align}
Using the fact that $I_x(b,a+1) = I_x(b,a) + \frac{x^b(1-x)^a}{aB(b,a)}$, we have
\begin{align}
E(\Delta T_1) = (E(T_1)-E(T_2))I_{1-x}(b,a) + t_{\text{pos}} \frac{R_1-R_2}{n+1}\frac{x^a(1-x)^b}{aB(a,b)}.
\end{align}
Similarly, the saved latency compared to an $(n,k_2,\ell_2)$ code is
\begin{align}
E( T_2 - T_{1,2}) = (E(T_2)-E(T_1))I_x(a,b) + t_{\text{pos}} \frac{R_1-R_2}{n+1}\frac{x^a(1-x)^b}{aB(a,b)}.
\end{align}

From \eqref{Latency of fixed code 1} and \eqref{Latency of fixed code 2} we can see that the latency of a fixed MDS code is a function of $n,R,\ell,t_{\text{pos}}$, and $t_{\text{trans}}$. One can optimize the code reconstruction threshold $R^\ast$ similar to \cite{lee2017speeding} and \cite{peng2020diversity} based on other parameters. However, the system parameters might change over time and one  ``optimal'' $R^\ast$ cannot provide low latency in all situations. For example, with fixed $n$, $\ell$ and the total information size, larger $t_{\text{trans}}$ results in a larger $R^\ast$ while larger $t_{\text{pos}}$ results in a smaller $R^\ast$. In our flexible codes, we can always pick the best $R_j$ over all $j \in [a]$, thus provide a lower latency.

\begin{figure}[!h]
\centering
\includegraphics[width=0.55\textwidth]{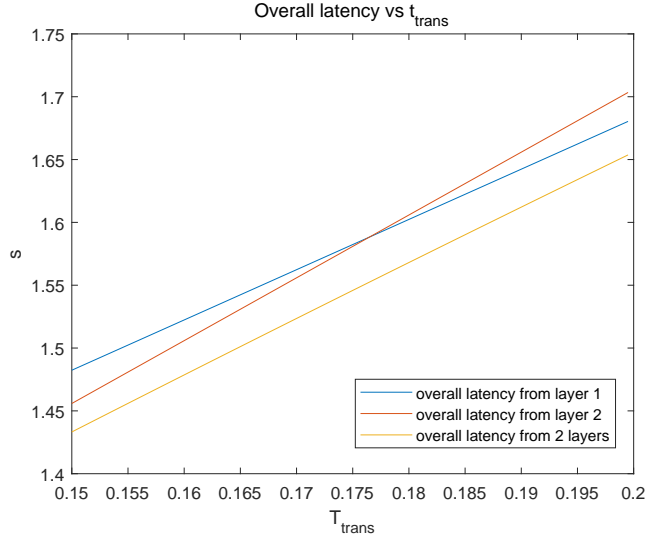}
\caption{Overall latency of fixed codes and flexible  codes. $n=16,R_1=15,R_2 = 12,\ell_1 = 4,\ell_2 = 5. t_{\text{pos}} = 1$.
%\bl{Change or add the example to be locality or PMDS. Say that our code provides a set of choices, instead of saying we are better in latency. Try to talk about complexity when the field size is twice more?}
}\label{HDD latency}
\end{figure}

Fig. \ref{HDD latency} shows the overall latency of fixed codes and flexible recoverable codes. We fix other parameters and change the unit data transfer time $t_{\text{trans}}$. For fixed codes, a smaller $R$ provides a lower latency with a smaller $t_{\text{trans}}$, and when $t_{\text{trans}}$ grows, a larger $R$ is preferred. However, our flexible code always provides a smaller latency, and can save $2\% \sim 5 \%$ compared to the better of the two fixed codes. 

\begin{figure}[!h]
\centering
\includegraphics[width=0.55\textwidth]{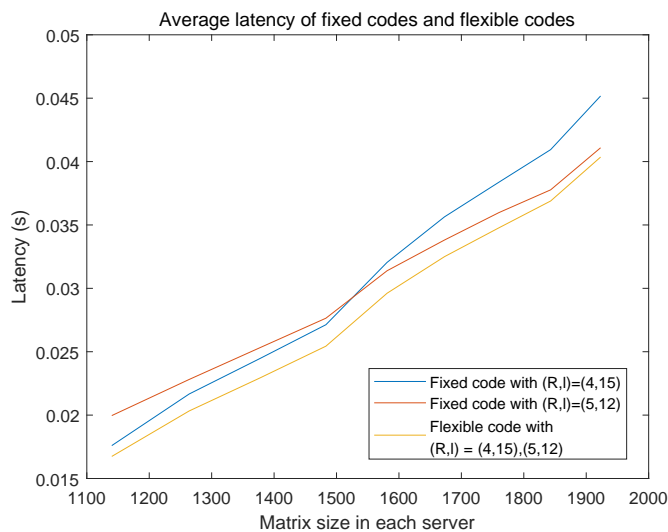}
\caption{Overall latency of fixed codes and flexible codes for matrix-vector multiplication in Amazon cluster. $n=8,R_1=5,R_2 = 4,\ell_1 = 12,\ell_2 = 15$.
%\bl{Change or add the example to be locality or PMDS. Say that our code provides a set of choices, instead of saying we are better in latency. Try to talk about complexity when the field size is twice more?}
}
\label{fig_amazon}
\end{figure}

Our flexible codes can also be applied to distributed computing systems for matrix-vector multiplications \cite{lee2017speeding}. The matrix is divided row-wisely and encoded to $n$ servers using our codes. Each server is assigned $\ell$ computation tasks. If any $R_j$ servers complete $\ell_j$ tasks, we can obtain the final results. Simulation is carried out on Amazon clusters with $n=8$ servers (m1.small instances). And each task is a multiplication of a square matrix and a vector. The results are shown in Fig. \ref{fig_amazon}. 
We can see a similar trend as that of Fig. \ref{HDD latency}. Our flexible code improves the latency by about 6\% compared to the better of the two fixed codes when the matrix size is $1500 \times 1500$.

\section{Conclusion}
\label{sec:conclusion}
In this paper, we proposed flexible storage codes and investigated the construction of such codes under various settings. Our analysis shows the benefit of our codes in terms of latency. Open problems include flexible codes for distributed computed problems other than matrix-vector multiplications, code constructions with a smaller finite field size and smaller sub-packetization, and storage codes utilizing partial data transmission from each node similar to universally decodable matrices.

\bibliographystyle{IEEEtran}
\bibliography{IEEEabrv,sample}
\end{document}